\documentclass{jfm}
\pdfoutput=1

\usepackage{graphicx}
\usepackage{xcolor}
\definecolor{myblu}{rgb}{0.1,0.1,0.5}
\definecolor{mygreen}{rgb}{0.13,0.26,0.26}
\usepackage{newtxtext}
\usepackage{newtxmath}
\usepackage{natbib}
\usepackage{hyperref}
\usepackage[acronym]{glossaries}
\usepackage{subfig}
\usepackage{ulem}
\hypersetup{
    colorlinks = true,
    urlcolor   = blue,
    citecolor  = black,
    linkcolor  = black
}

\newcommand{\RomanNumeralCaps}[1]
\linenumbers

\title{
Unsteadiness characterisation of 
shock wave/turbulent boundary-layer interaction 
at moderate 
Reynolds number
}

\author{Matteo Bernardini\aff{1}, 
  Giacomo Della Posta\aff{1}, 
  \corresp{\email{giacomo.dellaposta@uniroma1.it}}, 
  Francesco Salvadore\aff{2},
 \and Emanuele Martelli\aff{3}}

\affiliation{\aff{1}Sapienza University of Rome, Department of Mechanical and Aerospace Engineering, via Eudossiana 18, 00184, Rome, Italy
\aff{2} HPC Department, CINECA, via dei Tizii 6/B, 00185, Rome, Italy
\aff{3} Department of Engineering, University of Campania ``L. Vanvitelli'', Via Roma 29, 81031, Aversa, Italy}

\newacronym{dns}{DNS}{Direct Numerical Simulation}
\newacronym{les}{LES}{Large-Eddy Simulation}
\newacronym{stbli}{STBLI}{Shock Wave/Turbulent Boundary-Layer Interaction}
\newacronym{rms}{rms}{root mean square}
\newacronym{psd}{PSD}{Power Spectral Density}
\newacronym{wff}{WFF}{Wavelet Flatness Factor}
\newacronym{dmd}{DMD}{Dynamic Mode Decomposition}
\newacronym{weno}{WENO}{Weighted Essentially Non-Oscillatory}
\newacronym{df}{DF}{Digital Filtering}
\newacronym{pdf}{pdf}{Probability Density Function}
\newacronym{streams}{STREAmS}{Supersonic TuRbulEnt Accelerated navier-stokes Solver}
\newacronym{piv}{PIV}{Particle Image Velocimetry}
\newacronym{gpu}{GPU}{Graphics Processing Unit}

\begin{document}
\maketitle

\begin{abstract}
A direct numerical simulation of an oblique shock wave impinging on a turbulent boundary layer at Mach number 2.28 is carried out at moderate Reynolds number, 
simulating flow conditions similar to those of the experiment by \cite{dupont2006space}.
The low-frequency shock unsteadiness, whose characteristics have been the focus of considerable 
research efforts, is here investigated via the Morlet wavelet transform. 
Owing to its compact support in both physical and Fourier spaces, 
the wavelet transformation makes it possible to track the time evolution 
of the various scales of the wall-pressure fluctuations. 
This property also makes it possible 
to define a local intermittency measure, representing a frequency-dependent flatness factor, 
to pinpoint the bursts of energy that characterise the shock intermittency scale by scale. 
As a major result, wavelet decomposition shows that the broadband shock movement is actually 
the result of a collection of sparse events in time, each characterised by its own temporal scale. 
This feature is hidden by the classical Fourier analysis, which can only show the time-averaged behaviour. 
Then, we propose a procedure to process any relevant time series, 
such as the time history of the wall-pressure or that of the 
separation bubble extent,
in which we use a condition based on the local intermittency measure 
to filter out the turbulent content 
in the proximity of the shock foot and to isolate only the intermittent component of the signal.
In addition, wavelet analysis reveals the intermittent 
behaviour also of the breathing motion of the recirculation bubble behind the reflected shock, 
and allows us to detect a direct, partial correspondence between the most significant intermittent events 
of the separation region and those of the wall-pressure at the foot of the shock.
\end{abstract}




\section{Introduction}
\label{intro}

The interaction between shock waves and turbulent boundary layers represents a
major challenge for modern aerospace research, given its occurrence in a broad range of
engineering applications involving transonic, supersonic, and hypersonic systems.
Typical examples include high-speed intakes, over-expanded rocket nozzles,
transonic airfoil buffeting, and aerodynamics of high-speed vehicles or space launchers.
\glspl{stbli} must carefully be taken into account in the design process,
since they have the potential of harmfully impacting the performance of aerospace systems by
enhancing aerodynamic drag and heat transfer at the wall. The adverse pressure gradient imposed by the shock
can dramatically alter the structure of the boundary layer, inducing its thickening and, in the case of strong interactions,
large-scale separation. The latter is usually associated with concurrent complex phenomena as amplification of turbulence and noise,
formation of large-scale vortical structures, and unsteadiness involving a wide range of spatial and temporal scales.
Due to their relevance from the technological and scientific standpoint, \glspl{stbli} have been an active research field for more
than five decades, as demonstrated by the numerous review articles about the topic 
\citep{green1970interactions, adamson1980analysis, 
delery1985shock, dolling2001fifty, gaitonde2015celebrating}.

One of the most gripping aspects of \gls{stbli} 
is the low-frequency unsteadiness of the system, 
which consists in the appearance of streamwise oscillations of the 
separation shock characterised by a broadband motion at frequencies well below the typical
time scales associated with fine-grained turbulence in the 
upstream boundary layer. It has been highlighted that
the shock motion exhibits unsteady features that are similar 
among various \glspl{stbli} in different 2D and 3D canonical geometries, 
such as compression ramps, reflected shocks, blunt and sharp fins.
Several experimental and numerical works have greatly contributed to shed
some light on the origin of this unsteadiness, 
for which different mechanisms have been 
proposed~\citep{erengil91,beresh02,humble09}, 
mainly divided into two categories. On the one hand, 
according to the upstream mechanisms, the source of the unsteadiness is
the presence of low-frequency flow structures in the incoming turbulent 
boundary layer \citep{andreopoulos87, ganapathisubramani2009low}; 
on the other hand, according to the 
downstream mechanisms, the unsteadiness is related to the breathing 
motion of the separation bubble behind the shock, which expands and contracts periodically \citep{touber2009large, piponniau2009simple}. 
Although a general consensus has still not been reached, 
it has been argued that both mechanisms always coexist 
in all \glspl{stbli}: the downstream mechanism dominates for
strongly separated flows, whereas a combined mechanism 
dominates for weakly separated flows \citep{clemens2009shock}.

In the last decade, several studies have exploited high-fidelity simulation data sets
to extract the relevant dynamical features of both compression ramps and oblique shock reflections
\citep{pirozzoli2010analysis, grilli2012analysis, nichols2017stability, martelli2020flow, bakulu2021jet}.
For example, \citet{priebe2016low} applied \gls{dmd} to the data obtained through 
\gls{dns} of a Mach 2.9 compression ramp \gls{stbli}, 
and found a strong similarity between \gls{dmd} modes and 
linear stability modes reported in the literature.
Furthermore, they revealed the existence of streamwise elongated regions of low and high momentum
extending from the shock foot in the downstream flow that the authors interpreted as reminiscent
of G{\"o}rtler-like vortices, already observed in previous studies \citep{loginov2006large}.
Similar findings were reported by \citet{pasquariello2017unsteady}, 
who performed a wall-resolved, long-time integrated \gls{les} of
a high-Reynolds number \gls{stbli} at Mach 3 with massive flow separation.
Their sparsity-promoting \gls{dmd} identified two types of modes: 
a low-frequency mode involving the main actors of the interaction (shock system,
separated shear-layer and separation bubble) associated with the 
classical breathing motion of the recirculating flow, and medium-frequency modes
associated with the shear-layer vortices produced at the shock foot and convected downstream. 

To better understand the unsteadiness of the \gls{stbli} system, 
time series extracted from the numerical or the experimental fluid domain
are usually analysed by means of the classical Fourier trasform. 
This standard technique is typically used to 
generate frequency spectra of the physical variables, 
to determine the energy content of the signals at each frequency, and 
has been applied extensively to characterise the motion of the reflected shock in \gls{stbli}, 
as reported in several works \citep{clemens2014low}. 
However, according to \citet{bell1995analysis}, 
such an approach is fundamentally justified only when stationary or periodic, 
ergodic data are measured over a long time. 
In fact, for highly unsteady and irregular time series, 
Fourier transform may misrepresent localised features of the signal
in a time-averaged sense, which would distort the 
representation of the actual physical phenomena taking place
\citep{huang1998empirical}.
The wavelet transform \citep{mallat1989theory, daubechies1992ten} is instead an analysis 
tool well-suited to the study of multi-scale and non-stationary processes. Indeed, 
decomposing a time series into the time/temporal-scale space 
makes it possible to detect the localised variations of the energy within the signal, 
through a localised counterpart of the standard Fourier spectra.  
As a result, wavelet transform allows the detection of intermittent or modulated features 
in complex flows. On the basis of such properties, wavelet analysis 
has provided valuable insight into fluid mechanics \citep{liandrat1990wavelet},
as attested by its use for the study of turbulence \citep{farge1992wavelet, farge1999turbulence}, 
aeroacoustics \citep{grizzi2012wavelet, camussi2017statistical}, 
transonic buffet \citep{kouchi2016wavelet}, and over-expanded nozzles \citep{martelli2017detached}.
However, only very few works have used it to characterise the low-frequency unsteadiness 
in \gls{stbli} \citep{poggie1997wavelet, guo2020strong}, 
and none of them performed such an analysis on a detailed \gls{dns} database.
Moreover, given the fact that intermittency is defined by the presence of 
localised bursts of high-energy activity, 
wavelets can represent a suitable basis also to map and estimate the importance of rare but strong events: 
a task that could be hardly fulfilled by the trigonometric functions used in the Fourier transform, 
because of their infinite support in the time domain.

In this work, we present novel \gls{dns} data of an oblique shock wave impinging on a turbulent boundary layer 
inspired to the experiment performed at the IUSTI's hypo-turbulent supersonic wind tunnel 
in \citet{dupont2006space, dupont2008investigation}. 
The flow we simulated is characterised by a Mach number equal to 2.28 and a
momentum-based Reynolds number $Re_\theta$ equal to 6900.
Several groups have previously studied the same flow case 
either at the same conditions of the experiment, but using 
\gls{les}~\citep{touber2009large, morgan13, agostini2015mechanism}, 
or using \gls{dns}, but at a reduced Reynolds number~\citep{pirozzoli2011direct},  
because of the hindering computational burden associated to the simulation.
%
%
Unlike \glspl{stbli} at low Reynolds numbers 
\citep{Dolling1985, adams2000direct, pasquariello2014large, nichols2017stability}, 
in which the viscosity diffuses the separation shock into a milder compression fan, 
the moderate Reynolds number considered allows the shock foot to penetrate 
more deeply into the boundary layer. As a result, the inviscid pressure jump 
of the oscillating shock leaves a direct footprint in the 
wall-pressure, which exhibits a clear and definite intermittency. 
For lower Reynolds number, instead, since the compression is less sharp, 
the transition between the pressure states before and after the shock is smoother, 
which results in an attenuated intermittency and a broader range of 
frequencies covered by the shock unsteadiness \citep{ringuette2009experimental}.
Thanks to the possibility of considering the direct trace of the shock unsteadiness at the wall, 
we are able to perform an exhaustive wavelet analysis
on the wall-pressure signal that demonstrates the power of a description of the \gls{stbli} unsteadiness
in the time/time-scale domain.
The continuous wavelet transform is first directly applied to the unsteady signal 
of the pressure at the wall, to identify the time evolution of the temporal scales characterising 
the wall-pressure signature of the present \gls{stbli}.
Then, we use a wavelet-based indicator to 
evaluate the degree of intermittency of the wall-pressure, and to 
extract from the signal its most energetic, intermittent part.
Inspired by \citet{poggie1997wavelet} and by \citet{camussi2021intermittent}, 
we propose to use the wavelet transform to filter the wall-pressure and to separate the
contributions from the flow turbulence and the separation shock unsteadiness.
Our work tries thus to exploit the compact support of the wavelet basis to 
highlight the intermittent nature of the large-scale unsteadiness of the system
and to define a practical procedure to extract relevant sparse features 
from signals of interest for the system modelling.
The development of similar procedures to process the pressure -- but also other quantities -- is a 
prerequisite for further investigations on \gls{stbli} systems.

Finally, we consider the dynamics of the separation bubble behind the separation shock.
According to aforementioned theoretical, experimental, and numerical works 
\citep{dupont2006space, piponniau2009simple, pasquariello2017unsteady}, the
motion of the shock system -- especially for strong interactions -- is strictly related 
to the breathing motion of the separated region. 
For this reason, we attempt to relate the intermittent features of the wall-pressure 
at the foot of the separation shock to the dynamics of the separation bubble, 
by replicating the wavelet intermittency analysis on a signal recording an estimate of the 
recirculation region extent on a single vertical slice throughout the 
duration of the simulation. Wavelet cross-spectra and wavelet coherence 
are also considered to estimate the relationship between the wall-pressure at 
the shock foot and the area of the separation bubble.

The paper is organised as follows: Section \ref{numset} presents the 
numerical setup of the simulations; Section \ref{database} describes the 
database generated and the validation carried out; Section \ref{results} 
presents the results of the analysis; finally, Section \ref{conclusions} 
reports some final comments.
%
%

\section{Numerical setup} \label{numset}

The results presented in this work are obtained from 
Direct Numerical Simulations performed using
\gls{streams} \citep{bernardini21}, a high-fidelity solver of the compressible Navier-Stokes equations,
targeted to canonical wall-bounded turbulent high-speed flows. The code, freely available online,\footnote{\url{https://github.com/matteobernardini/STREAmS}} 
solves the complete set of Navier-Stokes equations for a perfect, heat conducting gas.
%
%
%

The equations are discretised on a Cartesian mesh and solved by means of the finite difference approach. The convective terms
are discretised using a hybrid energy-conservative shock-capturing scheme in locally conservative form~\citep{pirozzoli10}.
In particular, a sixth-order, central, energy-preserving flux formulation is adopted in smooth regions of the flow, which ensures
a robust and accurate discretisation of the wide range of spatial and temporal scales typical of turbulence, 
without relying on numerical (artificial) diffusivity. Shock-capturing capabilities are achieved through the Lax--Friedrichs flux vector splitting,
where the characteristic fluxes are reconstructed at the interfaces using a fifth-order,
\gls{weno} reconstructions~\citep{jiang96}.
The hybridisation between the central and the \gls{weno} scheme is managed by the implementation of a shock sensor that computes the local smoothness
of the solution and identifies discontinuities into the flow. Sixth-order, central 
finite-difference approximations are also applied for
the discretisation of the viscous terms. The resulting system of ordinary differential equations
is integrated in time by means of a third-order, low-storage, Runge-Kutta scheme.
The solver is written in Fortran 90 and presents an MPI parallelisation based on a classical
domain-decomposition. The current version is able to run on NVIDIA multi-\glspl{gpu} architectures through the CUDA Fortran paradigm.
Additional details on the numerical methods and the scalability performance of \gls{streams} can be found in~\citet{bernardini21}.


\section{Database description and validation} \label{database}

The \gls{dns} database is inspired to 
the oblique shock-wave/turbulent boundary layer 
interaction reported in~\citet{dupont2006space, dupont2008investigation}, 
characterised by a free-stream Mach number $M_\infty = 2.28$, 
an incidence angle of the shock generator $\phi = 8^\circ$, 
and a Reynolds number $Re_\theta = 5100$. 
In the simulation, the Reynolds number based on the 
compressible momentum thickness of the incoming turbulent boundary layer
and on the freestream viscosity is equal to $Re_{\theta} \approx 6900$, 
which corresponds to $Re_{\delta_2} \approx 3900$ if wall viscosity is considered.  
The main properties of the upstream boundary layer are reported 
in table~\ref{tab:bl} for both the numerical and the experimental case.
\begin{table}
 \centering
\caption{Properties of the incoming boundary layer: experimental and numerical data.}
\label{tab:bl}      
\begin{tabular}{ccccccccc}
\hline\noalign{\smallskip}
	--  &  $M$ & $Re_{\theta}$ & $Re_{\delta_2}$ & $Re_{\tau}$ & $C_f$ & $\theta_{0,inc}/\delta_0$ & $H$ & $H_{inc}$\\
\noalign{\smallskip}\hline\noalign{\smallskip}
	Experiment & 2.28 & 5100 & 2915 & --   & $2.00 \cdot 10^{-3}$ & 0.11 & --   & --   \\
	DNS        & 2.28 & 6882 & 3900 & 1100 & $1.98 \cdot 10^{-3}$ & 0.10 & 3.21 & 1.33 \\
\noalign{\smallskip}\hline
\end{tabular}
\end{table}

The simulation is conducted in a computational domain of size $L_x/\delta_0 \times L_y/\delta_0 \times L_z/\delta_0 = 70 \times 12 \times 6.5$, 
the reference length $\delta_0$ being the thickness of the incoming boundary layer upstream of the interaction ($99 \%$ of the freestream velocity). 
According to results of \cite{morgan13}, 
the spanwise width of the computational domain that we use 
is sufficient to avoid any confinement effect and thus
any artificial increase in the size of the separation bubble.
The computational mesh consists of $N_x \times N_y \times N_z = 8192 \times 1024 \times 1024$ grid nodes, equally spaced in the wall-parallel directions.
In terms of wall units evaluated in the upstream boundary layer, the spacings in the streamwise and spanwise directions
are $\Delta x^+ \approx 7$, $\Delta z^+ \approx 5$.
A stretching function is applied in the wall-normal direction to increase the resolution in the near-wall region and in the interaction zone.
At the wall, the spacing $\Delta y^+$ varies in the streamwise direction and it ranges between $\Delta y^+ \approx 0.6$ 
and $\Delta y^+ \approx 1.1$, upstream and downstream of the interaction respectively. 

\begin{figure*}
     \centering
     \subfloat[]{
     \includegraphics[width=0.495\textwidth]{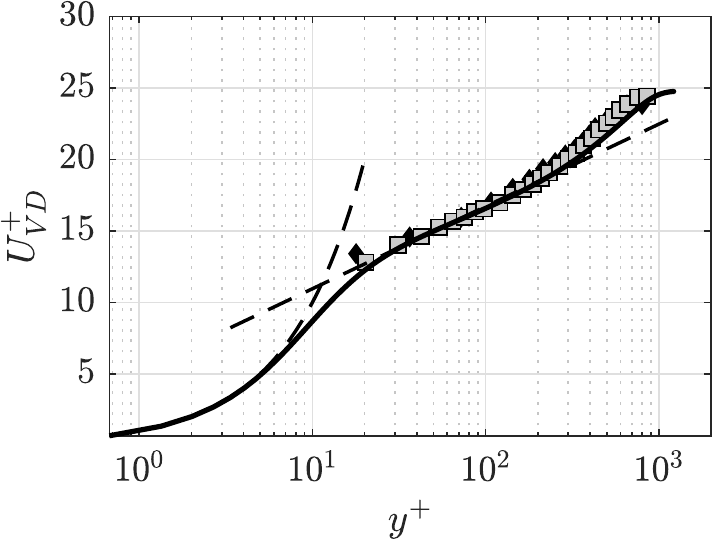}} 
     \subfloat[]{
     \includegraphics[width=0.495\textwidth]{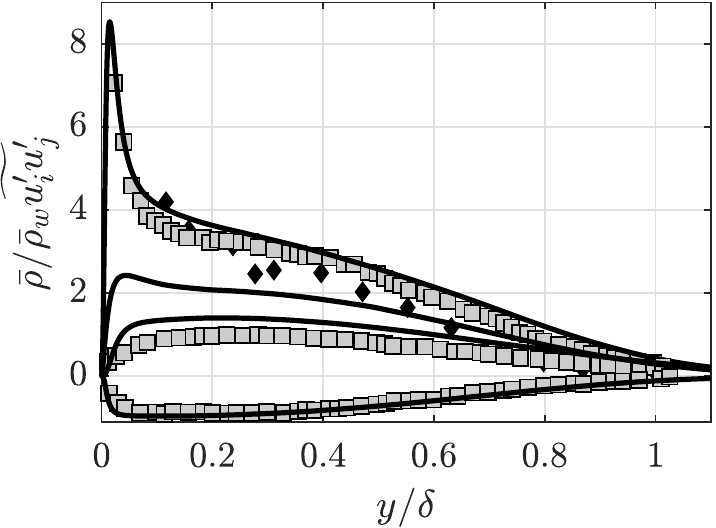}}
     \caption{Comparison of (a) van Driest–transformed mean velocity profile and (b) density-scaled Reynolds
        stress components for the incoming boundary layer with reference experimental data. Symbols denote
        experiments by \cite{elena1988experimental} (diamond, $M_\infty = 2.32$, $Re_\theta = 4700$) and 
        \cite{piponniau2009simple} (squares, $M_\infty = 2.28$, $Re_\theta = 5100$).}
     \label{fig:blay_check}
    \end{figure*}

The boundary conditions are specified as follows. At the outflow, non-reflecting
conditions are imposed by performing a characteristic decomposition in the direction normal to the boundary~\citep{poinsot92}.
A similar treatment is also applied at the top boundary away from the incoming shock, where instead the inviscid Rankine-Hugoniot
jump conditions are locally imposed to mimic the presence of a shock generator.
A characteristic wave decomposition is also employed at the bottom no-slip wall, where the wall temperature is set to
the recovery value for the upstream boundary layer.
A critical issue in the simulation of spatially evolving turbulent flows is the prescription of an inflow turbulence
generation method. In \gls{streams}, velocity fluctuations at the inlet plane are imposed by means of a synthetic \gls{df} 
approach~\citep{klein03}, extended to the compressible case by means of the strong Reynolds analogy~\citep{touber2009large}.
For \gls{stbli} computations, this type of approach is preferable with respect to alternatives based on the recycling-rescaling procedure~\citep{lund98}, 
because it does not introduce spurious frequencies that might interact with the dynamics of the reflected shock.
An efficient implementation of the method is here obtained using an optimised \gls{df} procedure~\citep{kempf12},
whereby the filtering operation is decomposed in a sequence of fast one-dimensional convolutions.
The implementation requires the specification of the Reynolds stress tensor at the inflow plane, which is interpolated by
a data set of previous \glspl{dns} of supersonic boundary layers performed by the same group~\citep{pirozzoli_url}.
Finally, in the spanwise direction the flow is assumed to be statistically homogeneous, and periodic
boundary conditions are applied.
In order to characterise the incoming turbulent boundary layer, 
figure \ref{fig:blay_check} shows a comparison of the van Driest–transformed mean velocity profiles
and of the density-scaled Reynolds stress components with the experimental data
from \cite{elena1988experimental} and \cite{piponniau2009simple}. 
A satisfying agreement can be observed for all the distributions, 
except for the wall-normal Reynolds stress component, 
which is typically underestimated by measurements. 

Various studies focused on the computational analysis of the confinement effects imposed by the presence of sidewalls
for \glspl{stbli} \citep{bermejo2014confinement, poggie2019flow, deshpande2021large}.
In particular, using wall-modeled \gls{les} with an equilibrium wall model
to replicate our same reference experiment, \citet{bermejo2014confinement} 
found that the mean separation bubble behind the interaction 
is characterised by a strong three-dimensionality imposed by the lateral confinement,
and observed that the size of the bubble is significantly larger than that predicted by spanwise-periodic simulations.
Therefore, to compare our results with the experimental measurements, in the following we use scaled interaction coordinates,
$x^* = (x-x_0)/L_{int}$ and $y^* = y/L_{int}$, where $L_{int}$ denotes the interaction length-scale, defined as the distance between the
nominal impingement location and the foot of the reflected shock, and $x_0$ denotes the mean position of the
ideal incident shock.\footnote{In the reference experimental work, $x_0$ is
equal to the mean position of the reflected shock. 
As a result, our values of $x^*$ 
are quantitatively different from those in \citet{dupont2006space}, 
even if the actual positions are 
correspondent.} 
For the present \gls{dns} data, the ratio $L_{int}/\delta_0$ is equal to $3.30$,
a value larger than that previously obtained at reduced Reynolds number ($L_{int}/\delta_0 = 2.89$) 
but smaller than the experimental measurement ($L_{int}/\delta_0 = 4.18$). 
Table \ref{tab:length_scales} reports a comparison of the length scales obtained 
with some references. 

\begin{table}
     \centering
    \caption{Comparison of the length scales of the \gls{stbli} considered ($M_\infty \approx 2.28$, $\phi = 8^\circ$) with some references.
    The superscript * indicates that the separation length has been estimated by the relation $L_{sep} = 0.8\,L_{int}$ \citep{clemens2009shock}.}
    \label{tab:length_scales}      
    \begin{tabular}{ccccc}
    \hline\noalign{\smallskip}
    	--  &  $M$ & $Re_{\theta}$ & $L_{int}/\delta_0$ & $L_{sep}/\delta_0$\\
    \noalign{\smallskip}\hline\noalign{\smallskip}
    	Experiment \citep{dupont2006space} & 2.28 & 5100 & 4.18 & 3.40\textsuperscript{*}  \\
    	Present case                       & 2.28 & 6882 & 3.30 & 2.16{\color{white}{\textsuperscript{*}}} \\
    	\cite{morgan13}                  & 2.28 & 4800 & 3.02 & 1.61{\color{white}{\textsuperscript{*}}} \\
    	\cite{aubard2013large}             & 2.25 & 3700 & 2.84 & 2.35{\color{white}{\textsuperscript{*}}} \\
    	\cite{pirozzoli2011direct}         & 2.28 & 2300 & 2.89 & 2.31\textsuperscript{*} \\
    	\cite{touber2009large}             & 2.30 & 5100 & 4.80 & 3.90{\color{white}{\textsuperscript{*}}} \\
    	\cite{pirozzoli2006direct}         & 2.25 & 3700 & 2.17 & 1.18{\color{white}{\textsuperscript{*}}} \\
     \noalign{\smallskip}\hline
    \end{tabular}
    \end{table}
    %

After an initial transient needed to develop the reflected shock and to achieve a statistically steady state,
the computation was advanced for a total time $T \approx 2000\,\delta_0/U_{\infty}$ which is much longer than
the previous \gls{dns} simulations. It is worth highlighting that the time interval here considered allows us to cover several
low-frequency cycles (approximately 25 cycles), thus making possible an in-depth analysis of the wall-pressure signature unsteadiness.
The simulation time step is $\Delta t \approx 0.0008\,\delta_0/U_{\infty}$ and the wall-pressure history
is recorded with a sampling time $\Delta t \approx 0.04\,\delta_0/U_{\infty}$.

\begin{figure}
 \centering
 \subfloat[$\bar{u}/U_\infty$  \label{fig:comparison_a}]{
 \includegraphics[height=0.95\textwidth,angle=270,clip]{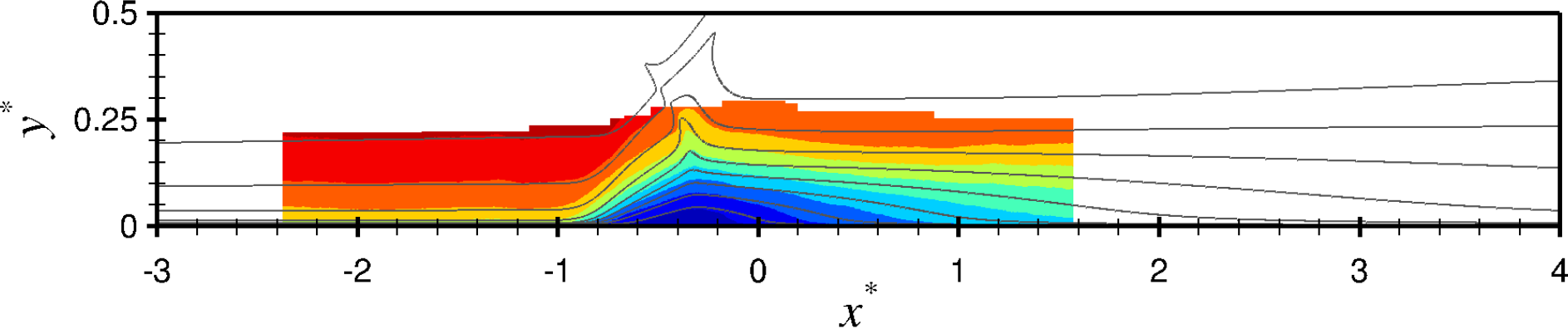}} \vskip 1em
 \subfloat[$\bar{v}/U_\infty$ \label{fig:comparison_b}]{
 \includegraphics[height=0.95\textwidth,angle=270,clip]{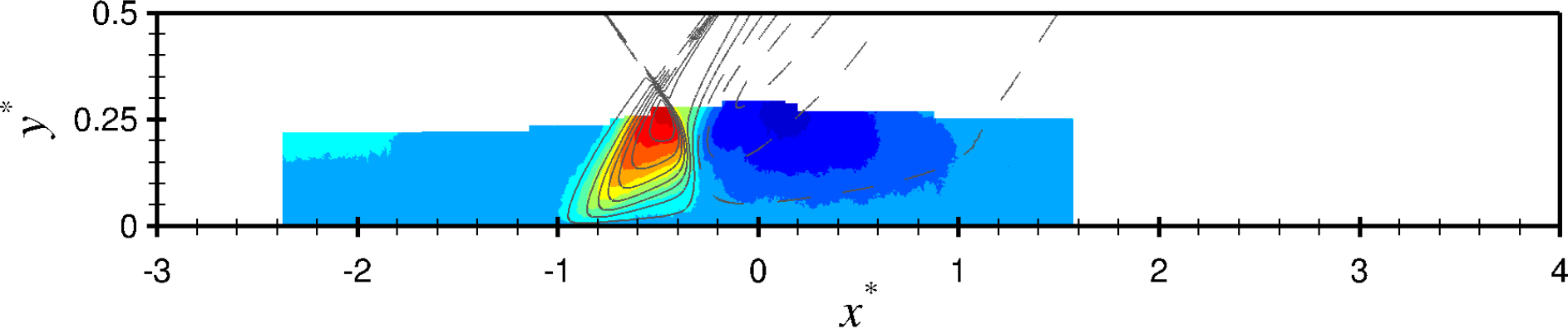}} \vskip 1em
 \subfloat[$\overline{u'^2}/U_\infty^2$ \label{fig:comparison_c}]{
 \includegraphics[height=0.95\textwidth,angle=270,clip]{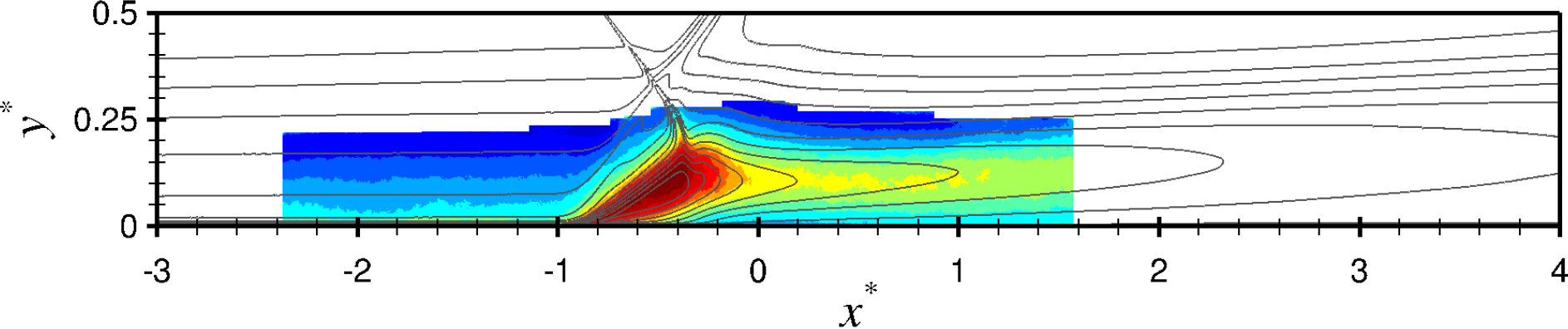}} \vskip 1em
 \subfloat[$\overline{v'^2}/U_\infty^2$ \label{fig:comparison_d}]{
 \includegraphics[height=0.95\textwidth,angle=270,clip]{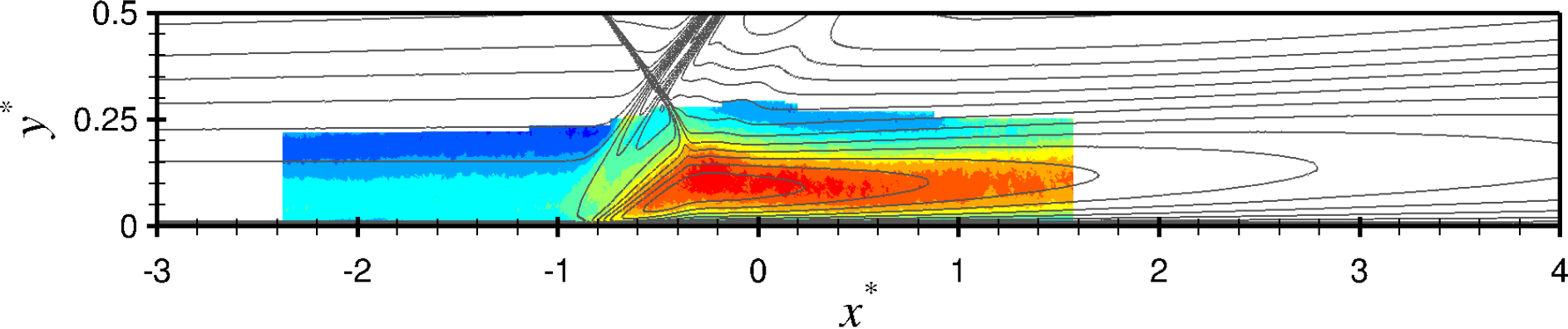}} \vskip 1em
 \subfloat[$-\overline{u'v'}/U_\infty^2$ \label{fig:comparison_e}]{
 \includegraphics[height=0.95\textwidth,angle=270,clip]{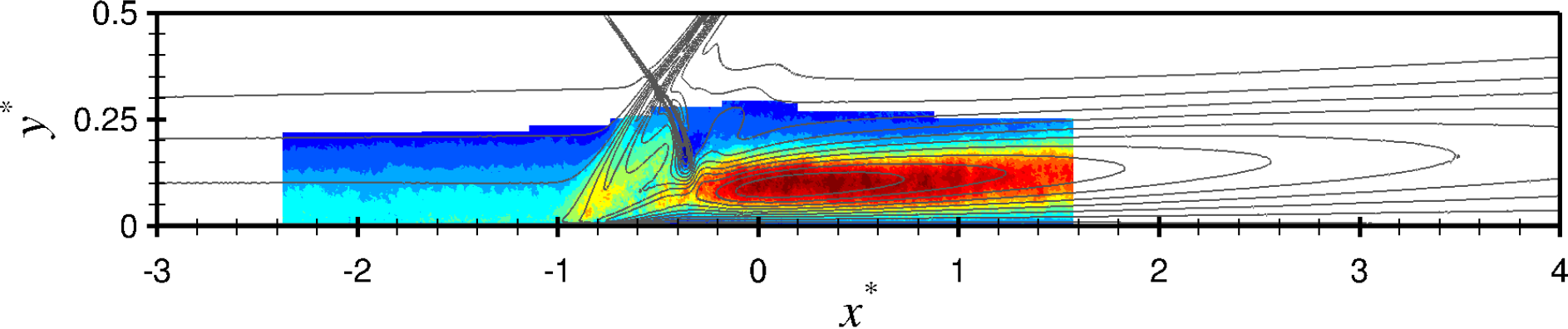}} \vskip 1em
 \caption{Comparison of velocity and Reynolds stress components on the spanwise centre plane in the interaction region. Colour maps refer to experimental PIV data, whereas DNS data
	are shown using contour lines. From top to bottom: mean streamwise velocity, mean vertical velocity, streamwise velocity fluctuations, vertical velocity fluctuations and Reynolds shear stress.}
 \label{fig:comparison}
\end{figure}

\begin{figure}
 \centering
 \includegraphics[width=\textwidth]{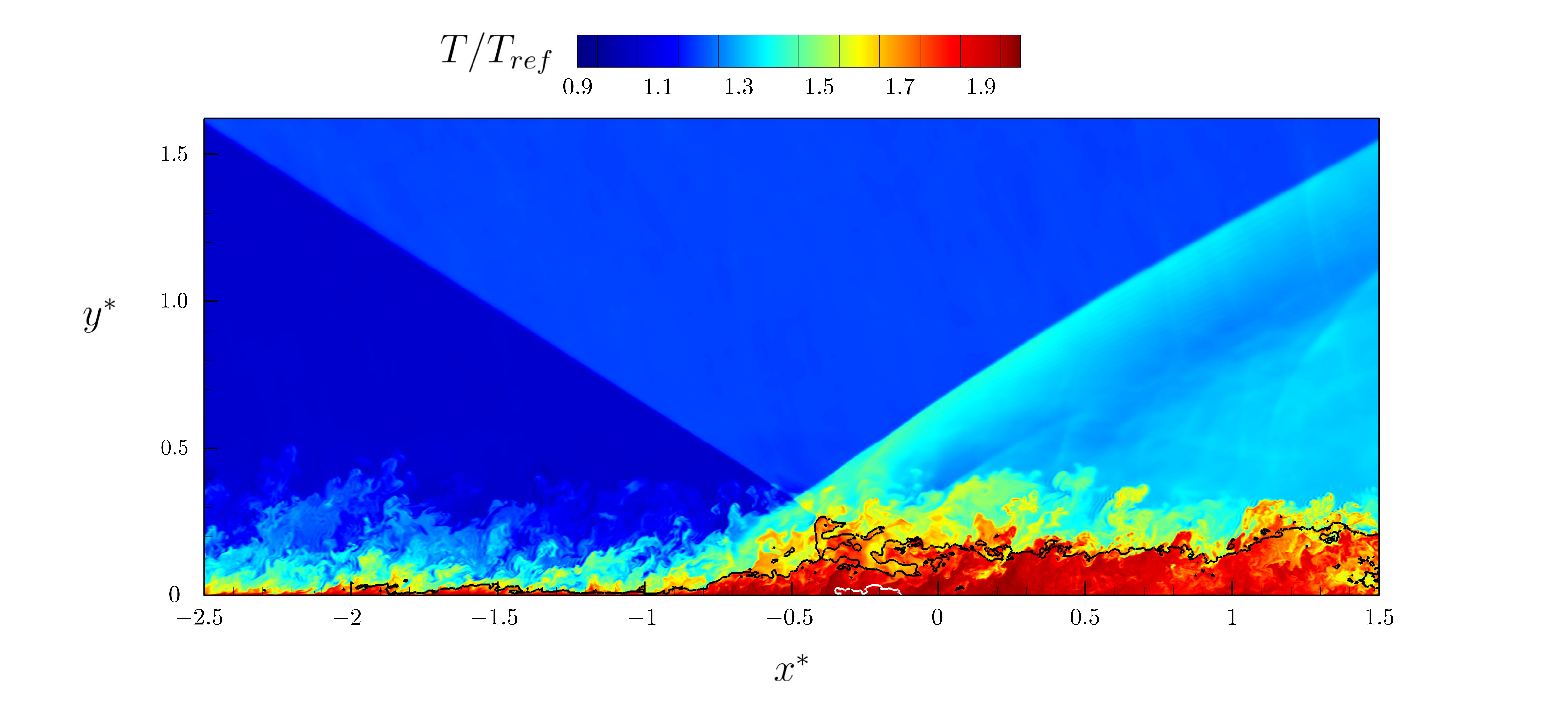} 
 \caption{Instantaneous contour of the non-dimensional temperature on a vertical slice. 
          The black line indicates the points characterised by Mach number equal to 1, 
          whereas the white line highlights the edge of the instantaneous recirculation bubble.}
 \label{fig:temperature_mach1}
\end{figure}

A comparison of the \gls{dns} results with the \gls{piv} data measured in the reference experiment is shown in figure~\ref{fig:comparison}, 
which reports the contours of the mean streamwise velocity, mean vertical velocity, streamwise velocity fluctuations,
vertical velocity fluctuations, and Reynolds shear stress. In the figure, contour lines obtained from \gls{dns}
are superposed onto heat maps derived from \gls{piv} in a limited region close to the interaction zone.
The comparison highlights that the overall structure of the interaction is recovered effectively. The numerical simulation reproduces correctly the thickening of the upstream boundary 
layer and the amplification of the turbulence across the interaction, 
which are associated with the shedding of vortices in the shear layer 
developing at the separation shock.
It is worth highlighting that, although a similar qualitative comparison was also previously observed at reduced Reynolds number,
the present \gls{dns} results provide a better quantitative agreement with the experiment
with respect to previous works, 
in particular with regard to the intensity of the turbulent fluctuations. 
Figure \ref{fig:temperature_mach1} shows an instantaneous contour of the non-dimensional temperature on a 
vertical slice, in which the black line indicates the locus of the points characterised by the sonic Mach number.
The instantaneous separation region is highlighted by indicating the edge of the local separation bubble. 
From the figure, we can observe the classical structure of the \gls{stbli}.
The thickening of the boundary layer, induced by the adverse pressure gradient from the shock system, makes 
the sweeps and ejections of the fluid towards and from the wall more evident. 
Moreover, the reflected shock generates an upward deflection of the flow before the virtual 
impingement point of the incident shock, thus promoting separation 
between $x^* = -0.5$ and $x^* = 0.0$. 
On top of the recirculation region, the innermost part of the incident shock interacts with the 
vorticity of the supersonic boundary layer in a non-trivial way, inducing even subsonic regions 
close to the intersection point between the incident and the reflected shocks. 
Finally, the following expansion and recompression 
that bring back the fluid to the original mean direction are also evident, especially in the top part of the 
slice. 
%
%
\begin{figure}
 \centering
\subfloat[$\bar{p}_w/p_\infty$  \label{fig:wallpressure_a}]{
 \includegraphics[width=0.49\textwidth]{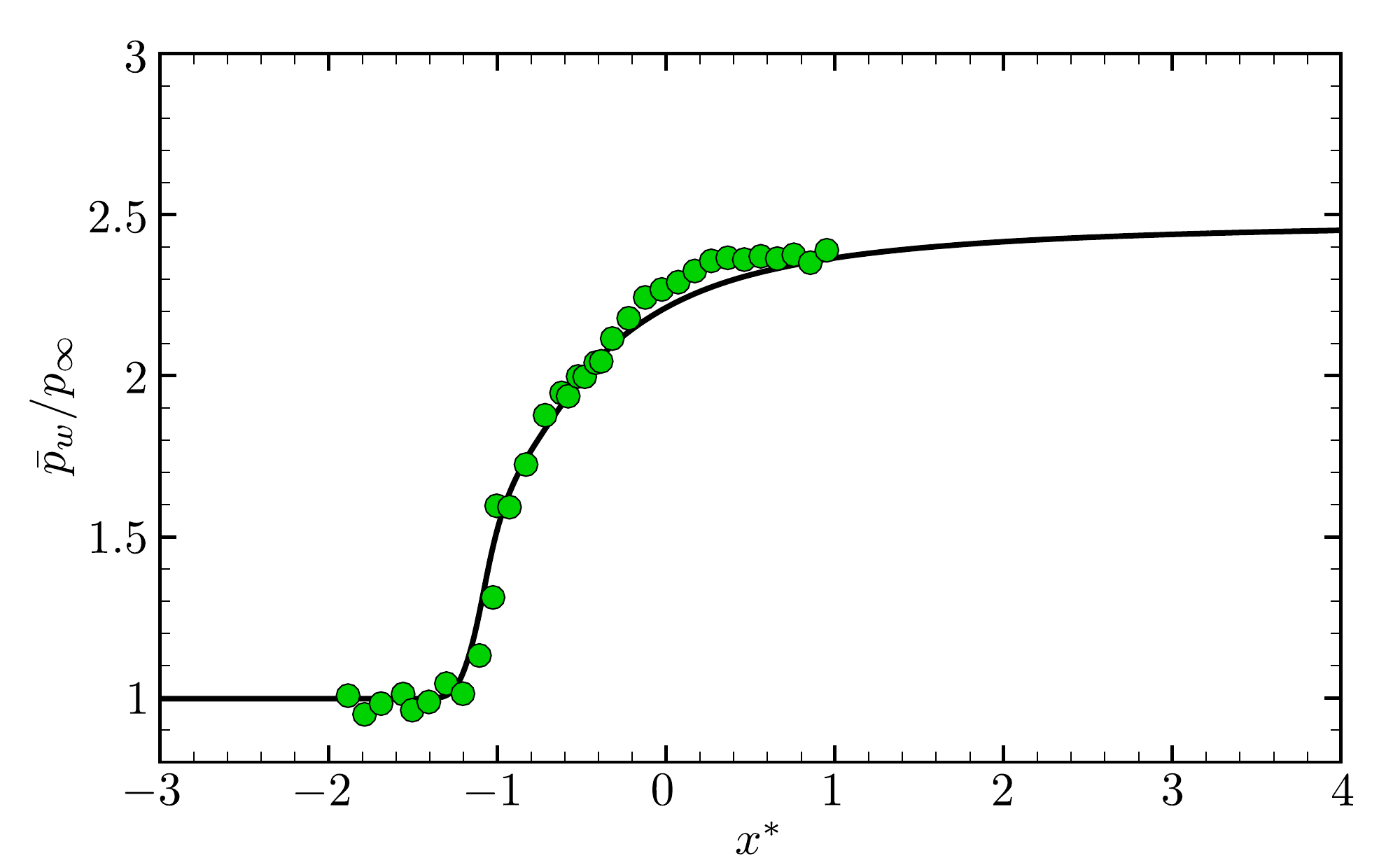}} 
\hfill
\subfloat[$\sqrt{\overline{p'^2}}/p_\infty$  \label{fig:wallpressure_b}]{
 \includegraphics[width=0.49\textwidth]{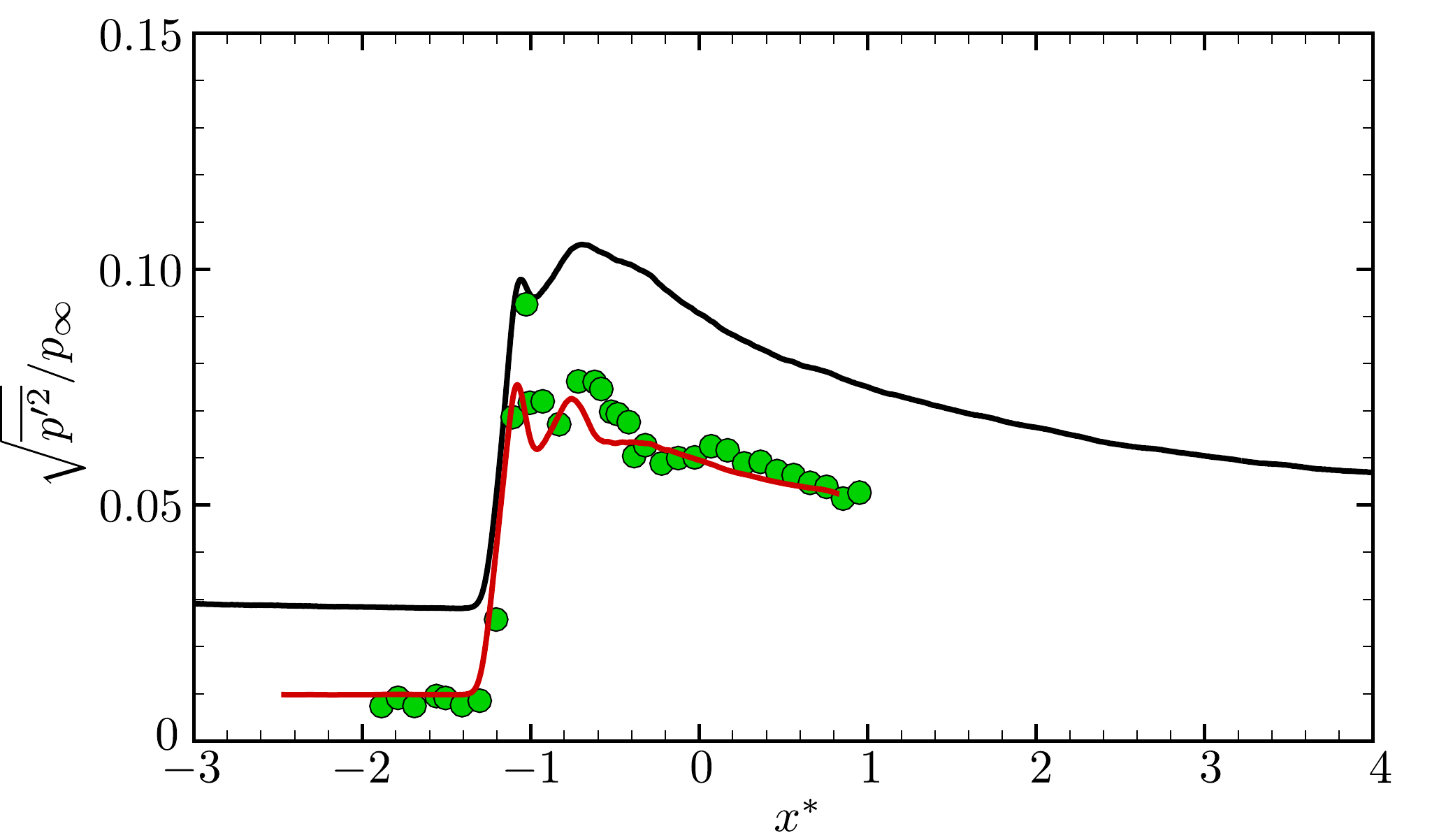}} 
	\caption{Streamwise evolution of (a) mean wall-pressure and (b) wall-pressure fluctuation intensity across the interaction region.
	{\color{green} $\bullet$} Experimental data \citep{dupont2006space}, -- Present DNS, {\color{red} --} 
	   Present DNS \gls{rms} evaluated by integrating the frequency spectra up to the cut-off of the experimental sensors.}
 \label{fig:wallpressure}
\end{figure}
%
%
\section{Analysis of the results} 
\label{results}
%
In this section, we focus on the behaviour of the wall-pressure, 
by investigating the frequency and the intermittent 
content of the time signals collected around the shock region
by means of classical Fourier analysis and by means of wavelet decomposition.
Extracting the features of this signal 
means understanding the behaviour of the shock unsteadiness, 
which is the main responsible for the dynamics of the wall-pressure fluctuations
besides turbulence.
In addition, we also provide an analysis of the dynamics 
of the separation bubble in the time/time-scale domain, 
and we discuss the relationship between the unsteadiness of the shock 
and the motion of the recirculating region.

\subsection{Time-averaged statistics of the wall-pressure}
At first, we focus on the behaviour of the mean pressure along the wall.
Figure~\ref{fig:wallpressure_a} shows the streamwise evolution of the mean wall-pressure, 
obtained by averaging the signal in time and in the spanwise direction, together with the experimental 
results from the reference work \citep{dupont2006space}.
Upstream of the interaction, the mean pressure is approximately constant and equal to the free-stream value. 
Later on, the pressure increases smoothly because of the presence of the oscillating shock, 
and after approximately $2\,L_{int}$ it reaches a plateau region. 
The comparison with the experimental data is satisfactory. 
Figure~\ref{fig:wallpressure_b} reports instead the comparison of the streamwise distribution of the 
\gls{rms} of the wall-pressure fluctuations. 
The typical behaviour shown in other \glspl{stbli} is observable.
At first, downstream of the foot of the shock, the intensity of the fluctuations increases suddenly
because of the backward and forward movement of the reflected shock. 
Then, in the separation and re-attachment region, the intensity decays, 
although the level of the fluctuations remains always higher than 
that of the attached boundary layer. 
The numerical results (solid black line) seems to show an overestimation of the fluctuation
intensity, but if the \gls{rms} is evaluated 
integrating the frequency spectra up to the cut-off 
of the experimental sensors (solid red line), 
the results agree well with the measurements.   

\begin{figure*}
 \centering
 \includegraphics[width=0.98\textwidth]{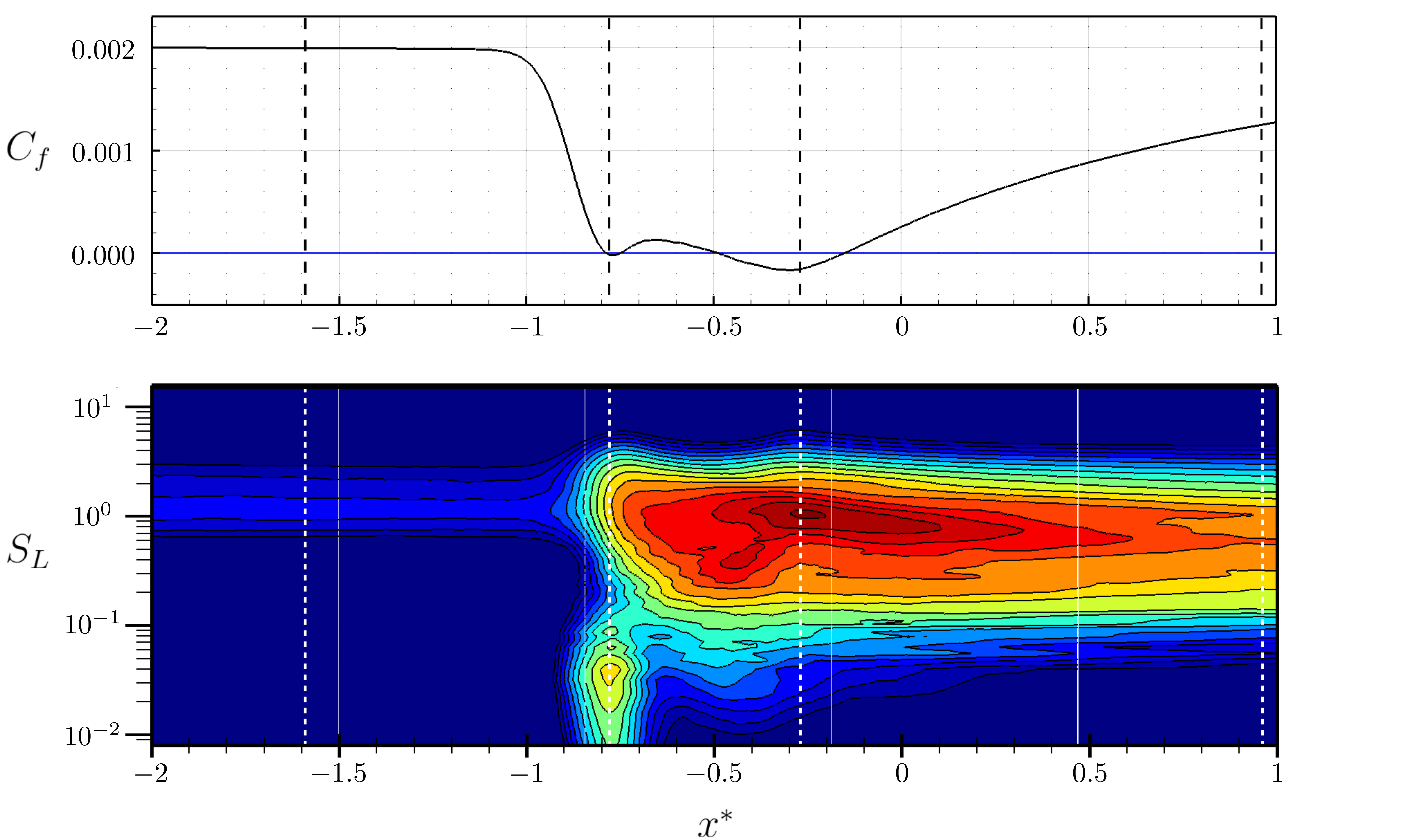} 
 \caption{Friction coefficient and premultiplied power spectral density of the wall-pressure along the interaction. The horizontal {\color{blue} blue} line
 in the first plot indicates the null value of $C_f$, whereas
 the vertical dashed lines indicate the position of the probes used for the sampling of the 
 wall-pressure signals.}
 \label{fig:mapspec}
\end{figure*}

\subsection{Fourier analysis of the wall-pressure}\label{subsec:fourier}
The premultiplied \gls{psd} of the wall-pressure signals has been computed for positions ranging 
from the zone before the shock to the 
relaxation zone, and the results are reported in figure~\ref{fig:mapspec}.
This map reveals the frequencies that contribute the most to the global energy of the local 
wall-pressure fluctuations, 
and the characteristics that clearly emerge agree with what was shown in~\citet{dupont2006space}.
In order to better interpret the different regions of the interaction, we report on top of the 
map also the distribution of the time- and spanwise-averaged friction coefficient $C_f = \tau_w / 1/2 \rho U_\infty^2$, where $\tau_w$ is the wall shear stress. 
Consistently with the literature, the mean friction coefficient is characterised by 
the presence of two negative peaks, where the upstream one has typically a larger amplitude in absolute value.
However, \cite{morgan13} showed that as the Reynolds number increases, the absolute magnitude of the 
first negative peak of the $C_f$ decreases, which thus explains the distribution observed in our case
leading to the appearance of two completely separate separation bubbles in the mean field.
However, the behaviour observed in the mean field is actually 
made of the superposition of different instantaneous states,
with the separation bubble alternating between a smaller and a 
significantly larger extent compared to what indicated by the negative 
mean friction coefficient.
On the basis of these observations, we define as a separation length  
the extent that goes from the first $C_f$ minimum at approximately 
$x^* \approx -0.8$ to the point with null $C_f$ at approximately $x^* \approx -0.15$. 
A separation length of $L_{sep}/L_{int} \approx 0.65$ or $L_{sep}/\delta_0 \approx 2.16$
is finally obtained, which is in line with the trend observed in \cite{morgan13} 
for increasing Reynolds number. 

\begin{figure*}
  \centering
 \includegraphics[width = 7cm]{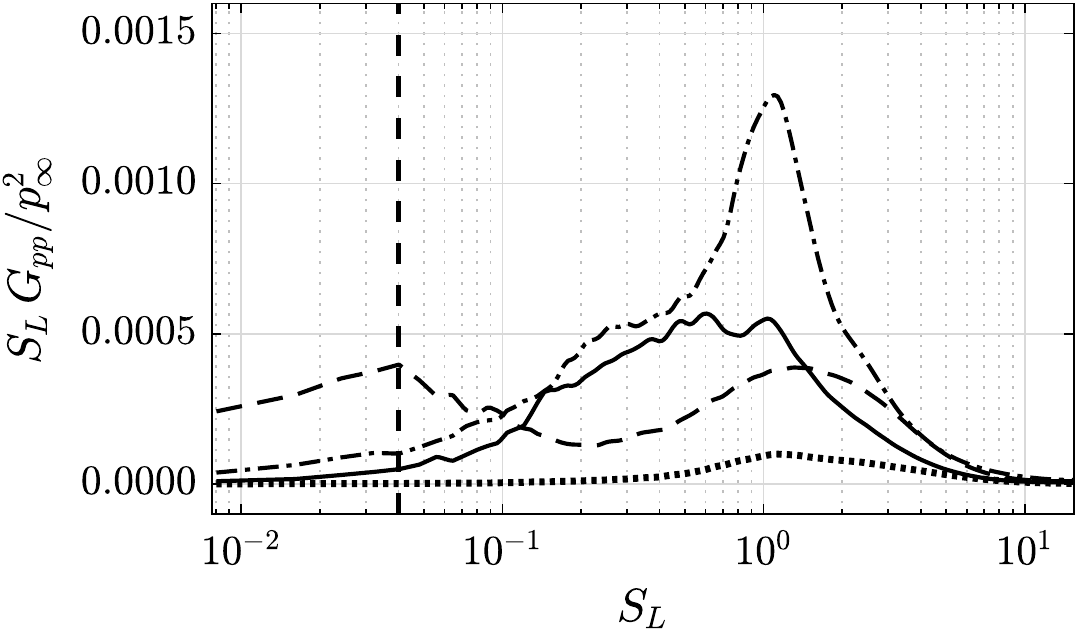} 
  \caption{Premultiplied power spectral density of the wall-pressure signal at:\\
  $\cdot \cdot \cdot$ $x^* = -1.60$, - - $x^* = -0.79$, -$\cdot$- $x^* = -0.27$, 
-- $x^* = 0.97$. 
           The vertical dashed line indicates the low-frequency peak at $S_L = 0.04$}. 
  \label{fig:fft_shock}
 \end{figure*}

The frequencies in the contour of figure~\ref{fig:mapspec} can be conveniently split into four separate zones, each involving its own characteristic temporal scales: 
\begin{enumerate}
\item the incoming turbulent boundary layer zone, $x^*< -0.8$, characterised by high-frequency contributions, with $S_L = f \, L_{int}/ U_\infty > 1$, and attached flow;  
\item the unsteady foot of the reflected shock, $x^* \approx -0.8$, characterised by low-frequency energy content.
The peak 
is widespread along a broad range of frequencies, suggesting 
an underlying shock unsteadiness more complicated than a simple periodic oscillation. 
In this region, the adverse pressure gradient imposed by the moving shock reduces the friction coefficient 
and forces the flow to start the separation;
\item the interaction zone, $-0.8 < x^*< -0.2$, characterised by intermediate turbulent scales developing in the detached supersonic shear layer. In this region, the spectrum reaches also its overall maximum in the point of minimal friction, 
which corresponds to the centre of the aft portion of the recirculation bubble;
\item the relaxation zone, $x^* > -0.2$, characterised by medium-frequency fluctuations developing in the separation region and dominating the boundary layer even after the reattaching point. 
\end{enumerate}
One of the most interesting characteristics revealed by the spectral map is the substantial distinction between the 
frequency content of the shock movement and  
the frequency content of the attached boundary layer, the separated shear layer, and the relaxation region.  
Figure~\ref{fig:fft_shock} shows the premultiplied \gls{psd} in correspondence of the highlighted probes 
at $x^* = -1.60$, $-0.79$, $-0.27$ and $0.97$ 
to better appreciate the spectral features of the fluctuations in points
representative of the above presented zones. 
In particular, at $x^* = -0.79$, the broad and high-amplitude peak 
at $S_L \approx 0.04$ 
reflects the low-frequency unsteadiness
of the shock motion, while the second wide bump with a peak at $S_L \approx 1.5$ 
is the trace of the incipient separated supersonic shear layer. 

%
\subsection{Wavelet analysis and detection of intermittency of the wall-pressure}
%
Elements of the wavelet theory can be found in several texts \citep{kaiser1994friendly, mallat1999wavelet}, 
whereas \citet{lewalle1994wavelet} and \citet{farge1992wavelet} discussed its applications to fluid mechanics. 
Here we report few elements to clarify the following discussion. 

The wavelet transform of a continuous signal $g(t)$ is defined as:
\begin{equation}
G_{\Psi}(k,\tau) = k \int_{-\infty}^{+\infty} g(t)\Psi^{*}(k(t-\tau))dt \,,
\label{eq:def}
\end{equation}
where $\Psi$ is the wavelet mother function, $k$ is a dilatation parameter,
$\tau$ is the time-translation parameter and $*$ denotes the complex conjugate.
By varying the wavelet scale $k$ and translating along the time with the time shift $\tau$,
one can construct a picture showing the amplitude of any feature at a certain scale and 
also its temporal evolution. 
In this study, the Morlet wavelet has been chosen since higher
resolution in frequency can be achieved when compared with other mother functions. 
An analytical expression for the complex Morlet wavelet is:
\begin{equation}
\Psi(t) = \pi^{-1/4} e^{i \omega_0 t} e^{-t^2/2} 
\label{eq:mother}
\end{equation}
where $\omega_0$ is the non-dimensional frequency, which is equal to 6 here to
satisfy the admissibility condition~\citep{farge1992wavelet}.
If $\hat{g}(\omega)$ is the Fourier transform of $g(t)$:
\begin{equation}
\hat{g}(\omega) = \int_{-\infty}^{+\infty} g(t) e^{-i \omega t} d\omega\,,
\label{eq:Fourier}
\end{equation}
then it follows that:
\begin{equation}
G_{\Psi}(k,\tau) =\frac{1}{2\pi} \int_{-\infty}^{+\infty} \hat{g}(\omega) \hat{\Psi}^*\left(\frac{\omega}{k}\right) e^{i \omega t} d\omega\,.
\label{eq:invFourier}
\end{equation}
As a consequence, the wavelet transform at a given scale $k$ can be interpreted as
a band-pass filter in the Fourier space.
Finally, following the method of \citet{meyers1993introduction}, the relationship between
the equivalent Fourier period and the wavelet scale can be derived
analytically by substituting a cosine wave of known frequency
into Equation~\ref{eq:Fourier} and computing the scale $k$ at which the wavelet power spectrum reaches its maximum.
More details about the operative algorithms adopted to compute the wavelet transform in this work 
can be found in \citet{torrence1998practical}.

In the following, we apply the wavelet analysis to the wall-pressure signal  
at four relevant stations, respectively at 
$x^* = -1.60$, $-0.79$, $-0.27$ and $0.97$ 
(see also figure~\ref{fig:mapspec} for the position of the numerical probes).
In particular, we first focus on the regions 
before and in the proximity of the foot of the reflected shock; 
then, we address the recirculating region, and finally we consider the re-attachment location. 
%
\begin{figure*}
  \centering
\subfloat{
\includegraphics[width = 0.49\textwidth]{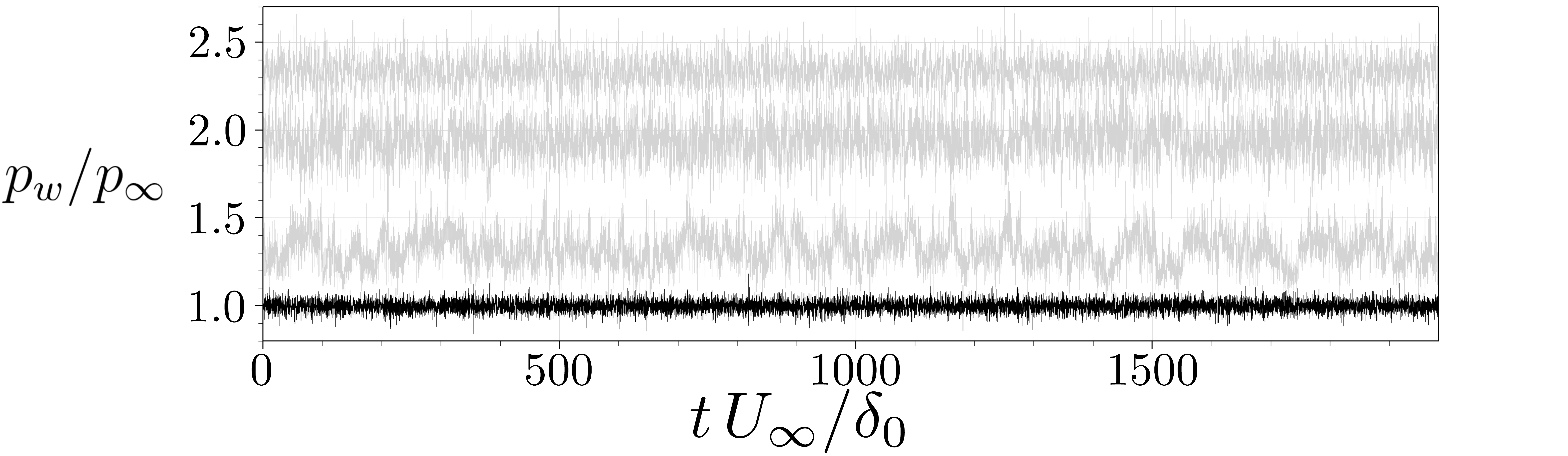}}
\hfill
\subfloat{
\includegraphics[width = 0.49\textwidth]{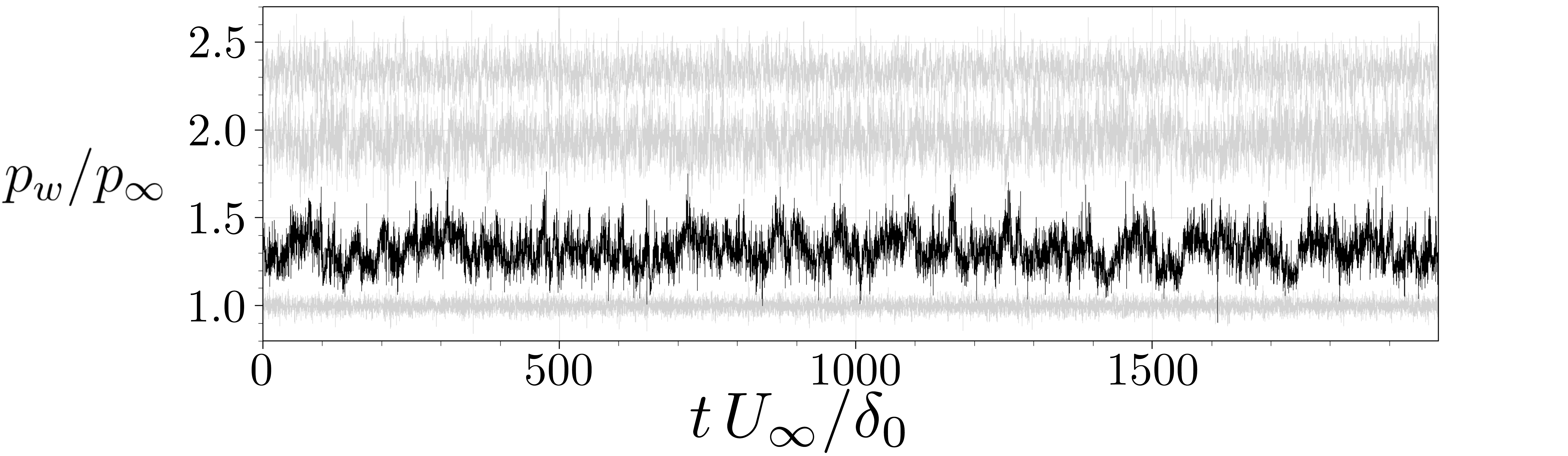}}\\[-2ex] 
\setcounter{subfigure}{0}
\subfloat[$x^* = -1.60$ \label{fig:spettri_160}]{
\includegraphics[width = 0.49\textwidth]{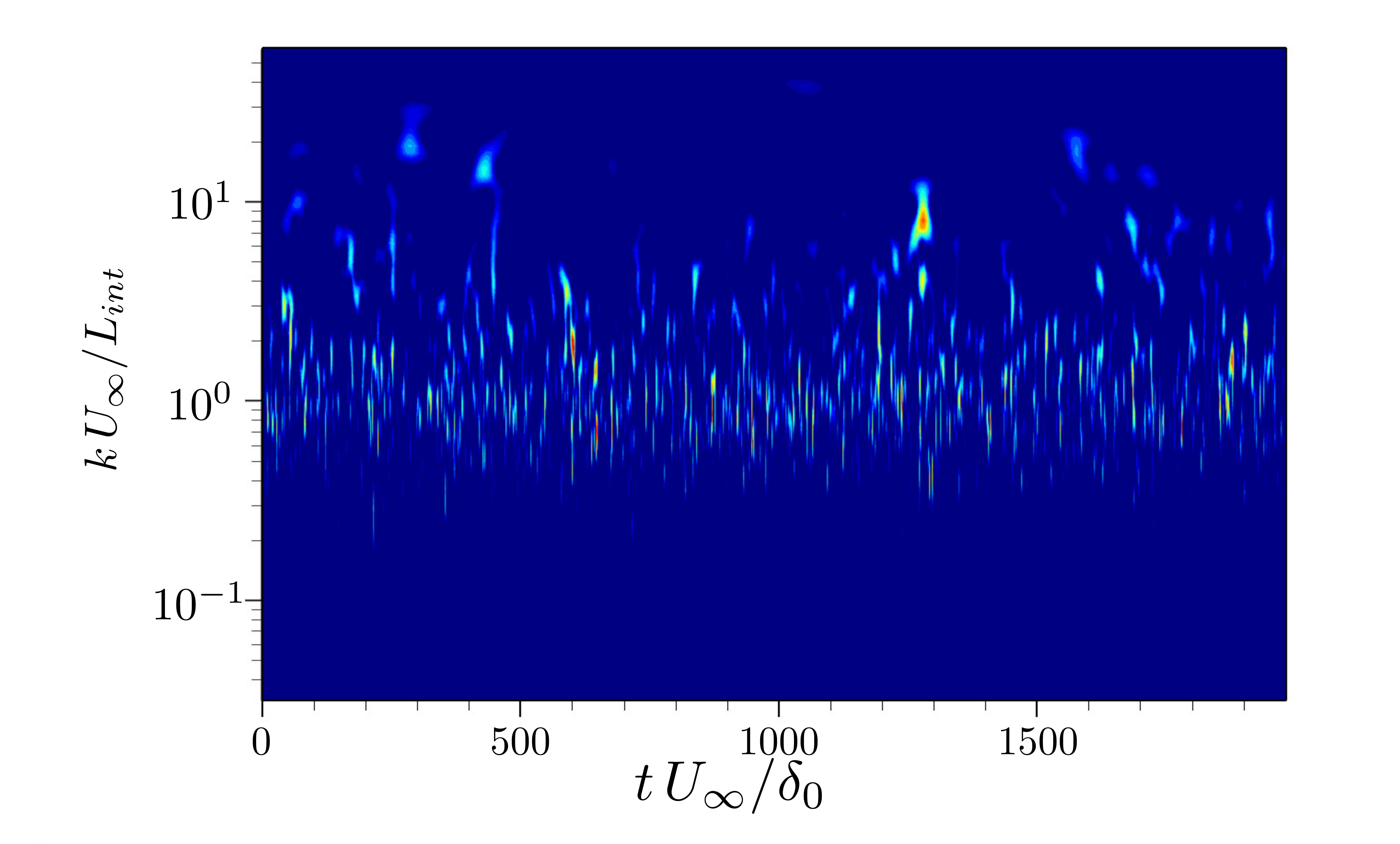}} 
\hfill
\subfloat[$x^* = -0.79$ \label{fig:spettri_079}]{
\includegraphics[width = 0.49\textwidth]{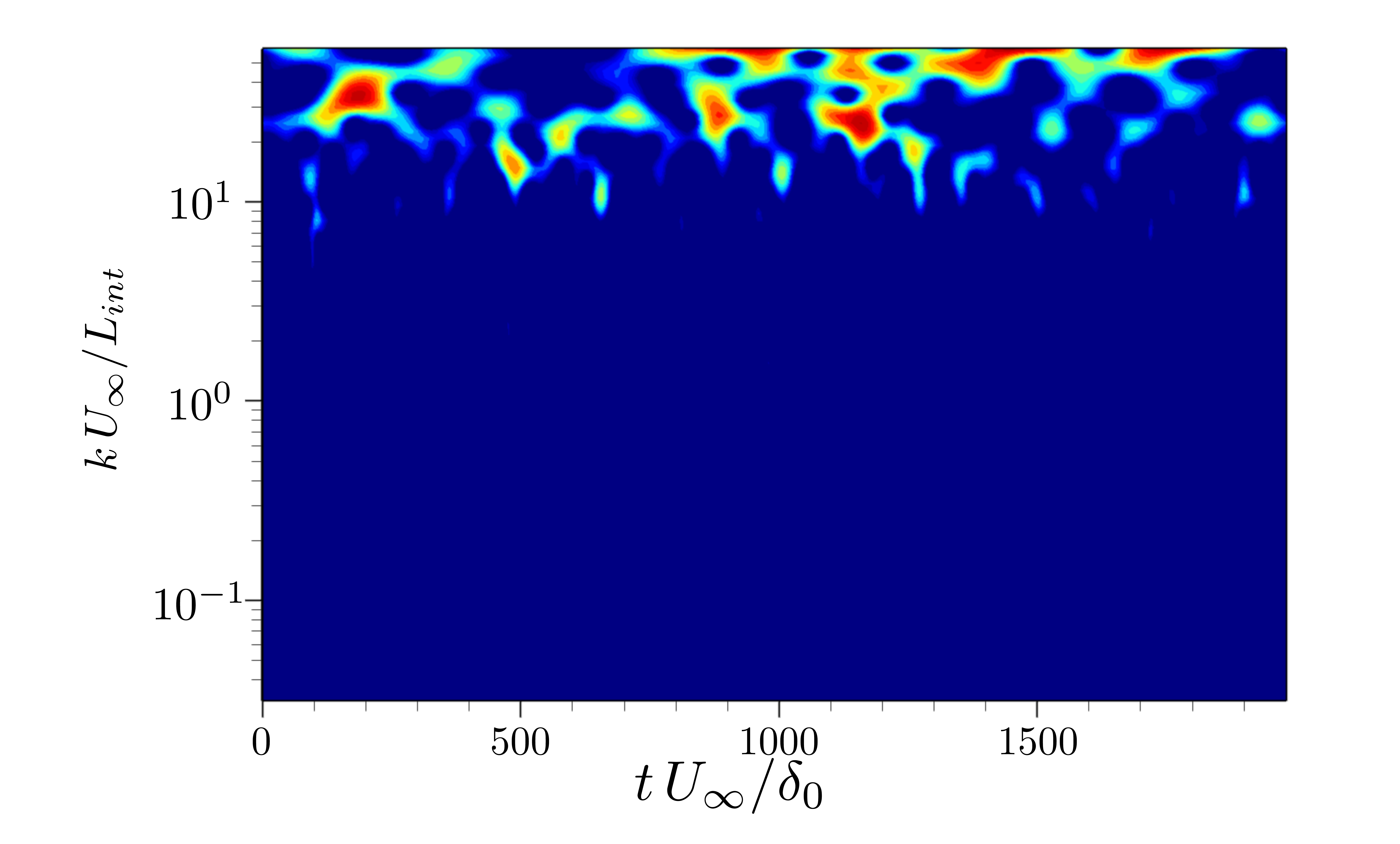}} 
\\
\subfloat{
\includegraphics[width = 0.49\textwidth]{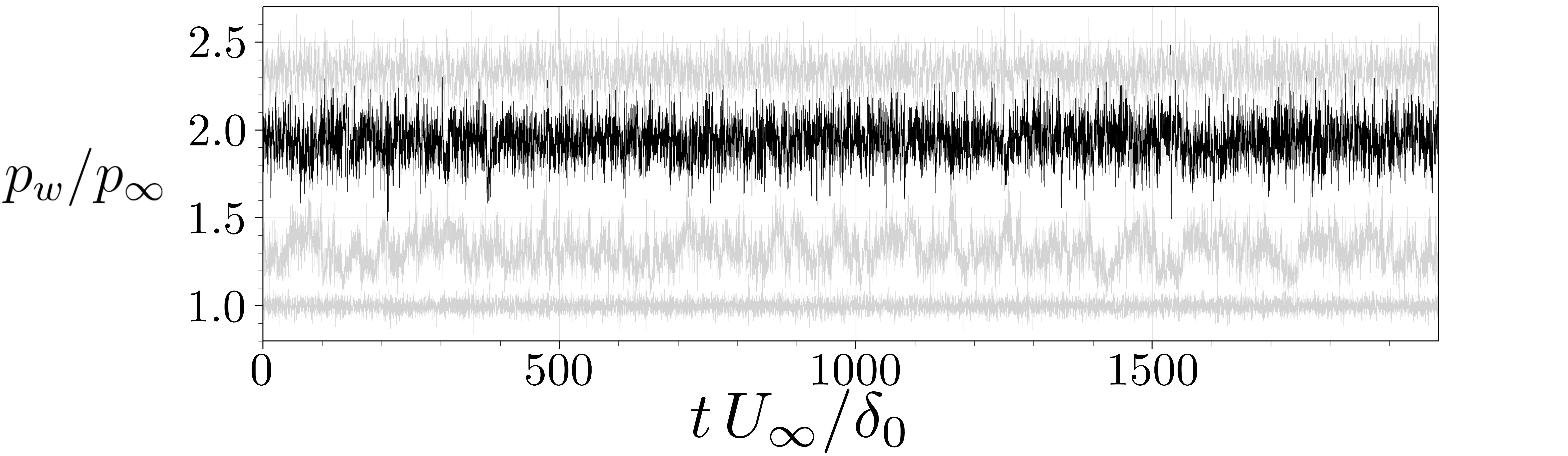}}
\hfill
\subfloat{
\includegraphics[width = 0.49\textwidth]{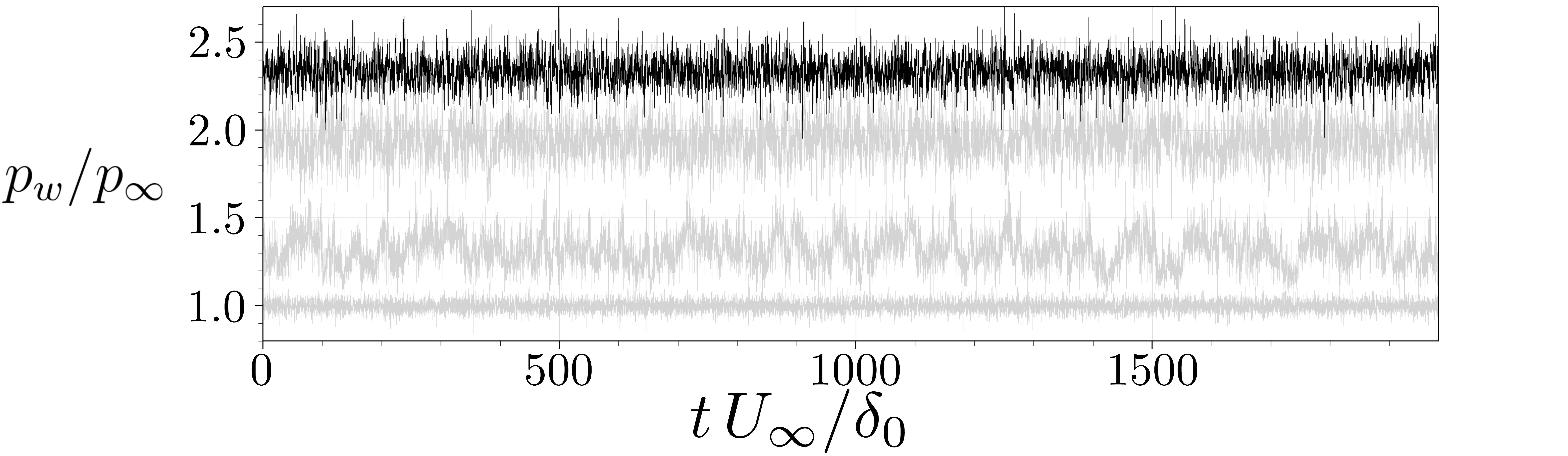}}\\[-2ex] 
\setcounter{subfigure}{2}
\subfloat[$x^* = -0.27$ \label{fig:spettri_027}]{
\includegraphics[width = 0.49\textwidth]{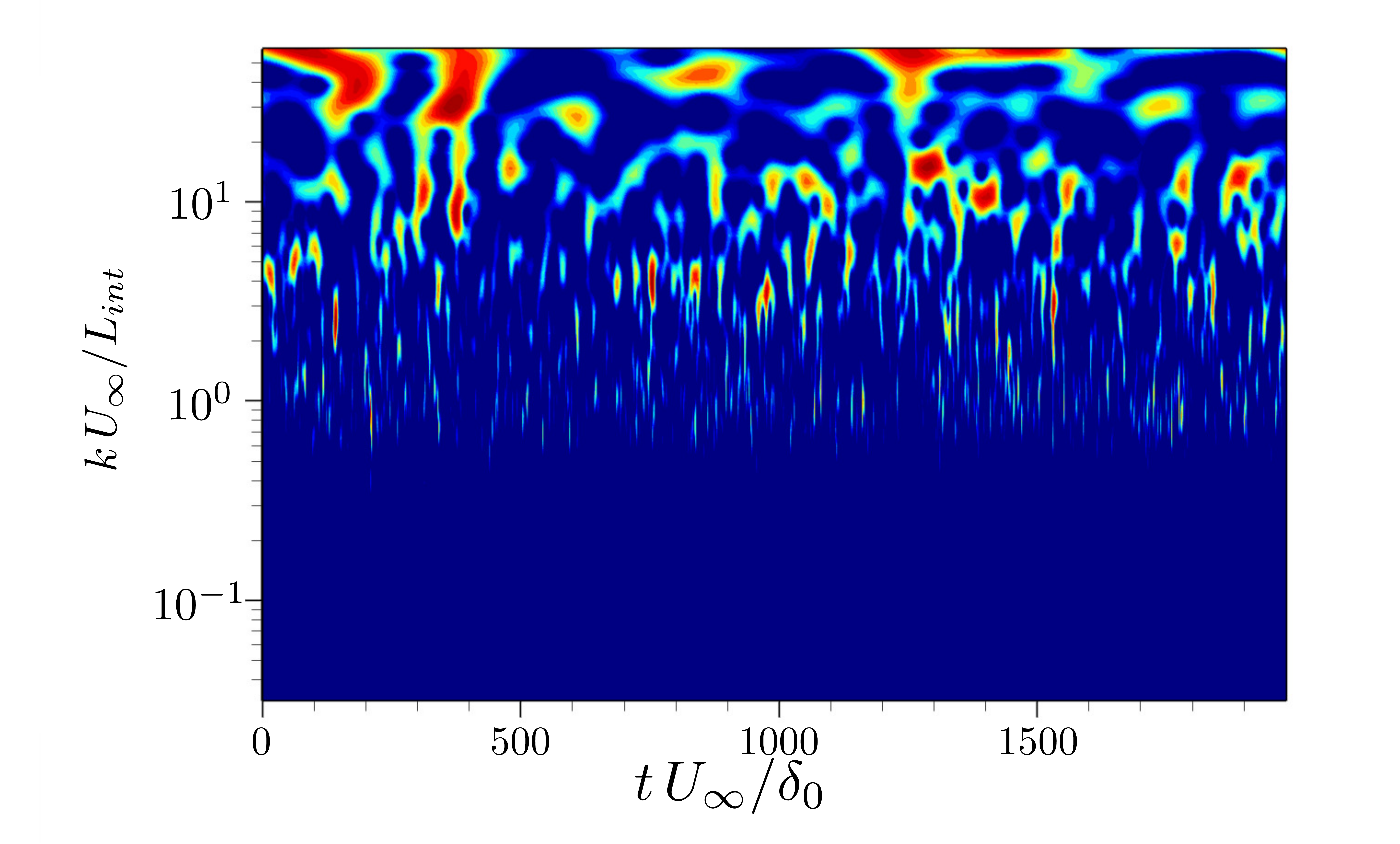}} 
\hfill
\subfloat[$x^* = 0.97$ \label{fig:spettri_097}]{
\includegraphics[width = 0.49\textwidth]{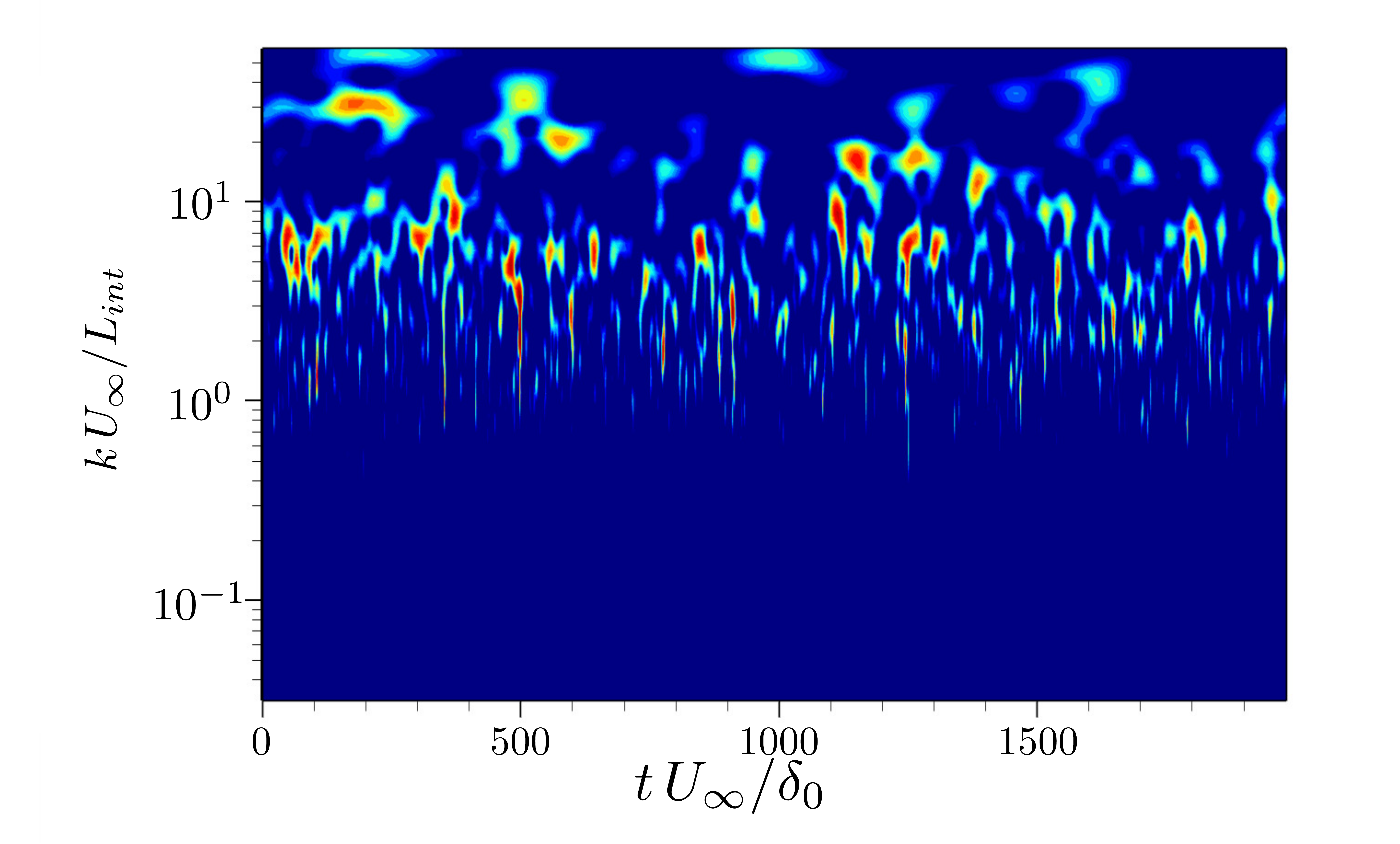}} 
  \caption{
           Wall-pressure signals (top) and
           Morlet wavelet transform modulus $|G_{\Psi}(k,\tau)|$ (bottom) at
          (a) $x^* = -1.60$, 
          (b) $x^* = -0.79$, 
          (c) $x^* = -0.27$, 
          (d) $x^* = 0.97$.
          For each probe, only values of the coefficients above 30\% of the relative maximum are shown.} 
  \label{fig:spettri}
 \end{figure*}
\begin{figure*}
  \centering
 \includegraphics[width = \textwidth]{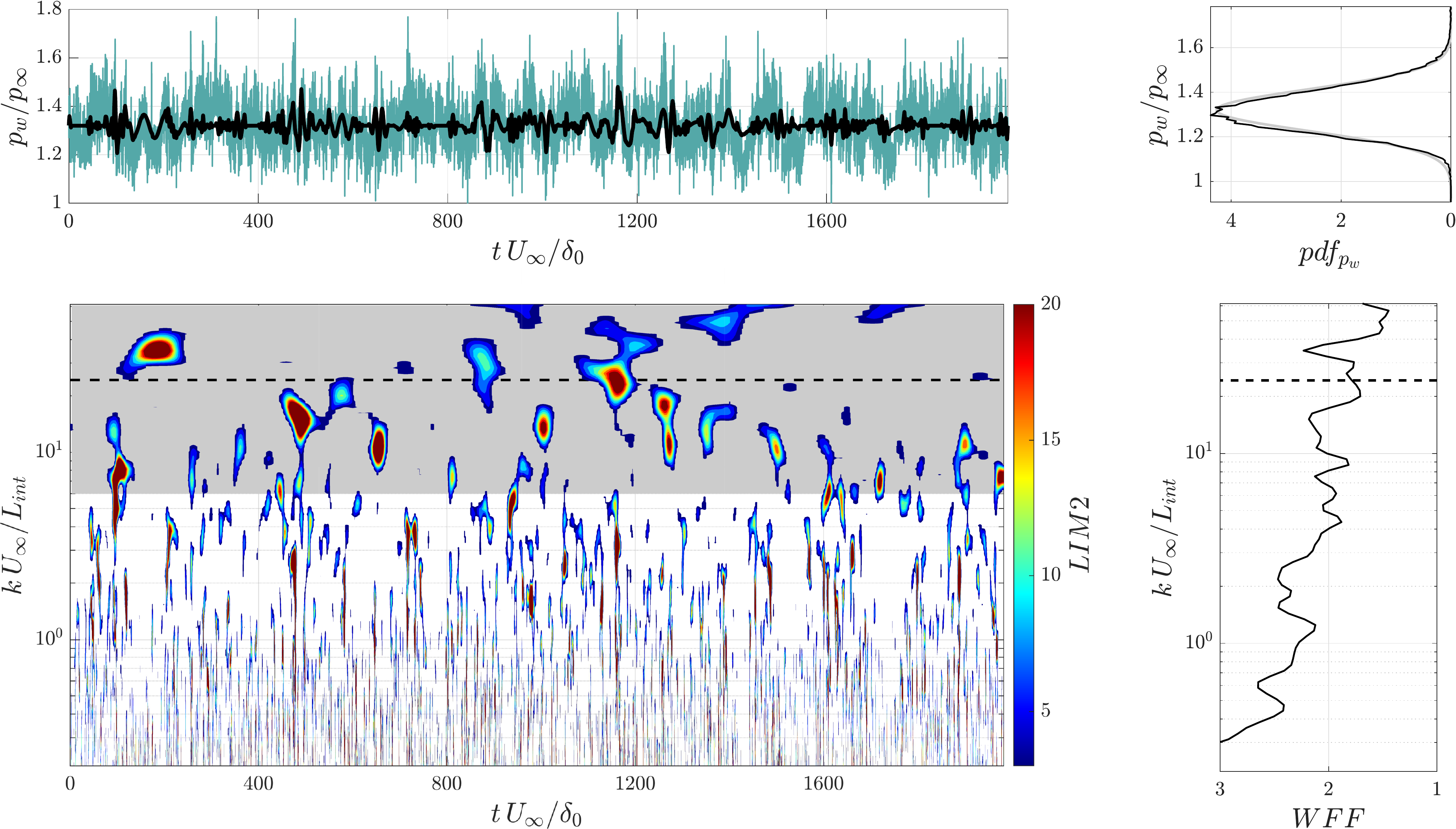} 
  \caption{
           On the top left, the {\color{mygreen}\textbf{green}} curve reports the original 
           wall-pressure signal at $x^* = -0.79$, 
           while the \textbf{black} curve reports its filtered-and-reconstructed part. 
           On the top right, the pdf of the original signal with 
           the corresponding fitted normal distribution in gray.
           On the bottom left, contour of $LIM2$ of $p_w/p_\infty$. 
           Values below 3 are cut off, whereas the gray area indicates the scales used for the filtering procedure. 
           On the bottom right, the corresponding WFF. 
           The horizontal dashed lines indicate the peak low frequency.}
  \label{fig:lim_1}
 \end{figure*}
Figure~\ref{fig:spettri} presents the modulus of the Morlet transform $G_{\Psi}(k,\tau)$,
the so-called scalogram, of the wall-pressure signals at 
the four prescribed locations.
The abscissa reports the non-dimensional time 
and the ordinate reports the temporal scale $k$ of the events in log-scale.
The blue color indicates a small magnitude of $|G_{\Psi}(k,\tau)|$, 
while the red color indicates a larger wavelet modulus, 
representing a more intense modulation of the amplitude 
of the pressure fluctuations within a certain range of temporal scales $k$ 
and at the given time instant $\tau$. Indeed, the real part of the wavelet coefficient 
is proportional to the amplitude of the fluctuation ($p'$), while the 
imaginary part is proportional to its time variation ($\partial p'/\partial t$) \citep{abe1999low}.
Figure~\ref{fig:spettri_160} shows the modulus of the Morlet transform for the wall-pressure 
recorded at $x^* = -1.60$, 
which is inside the boundary layer and before the foot of the reflected shock.  
It can be seen that, as expected, the 
attached boundary layer is characterised by small-scale 
eddies, which appear to be randomly distributed and fairly time-filling, 
although the spotted appearance underlines the intermittent character of such turbulent structures.   
Figure~\ref{fig:spettri_079} shows the wavelet modulus at $x^* = -0.79$, 
immediately downstream of the foot of the reflected shock. 
Given the position, the pressure signal from this probe is clearly modulated by the shock movement 
in the amplitude and in the characteristic temporal scales, 
or instantaneous frequencies if we consider the inverse of the temporal scales.
The contour confirms that the shock movement has a dominant content
at low frequencies/large time scales, as observed in the Fourier spectra. 
However, it reveals also that 
this motion is actually made of a collection 
of events localised in time and characterised by different temporal scales, 
larger than and well separated from the ones present in the upstream boundary layer. 
The spectral content from the Fourier analysis is thus the result of
an average in time of such events, whose combined effect is the broad low-frequency peak
spotted in figure \ref{fig:mapspec}. 
Figure~\ref{fig:spettri_027} characterises instead the pressure signal at $x^* = -0.27$, 
in the small separated region surrounded by the detached supersonic shear layer.
Here, the wavelet map is representative of a shock-induced turbulent recirculating flow:
the shock footprint is well visible from the distribution in time of large scale events, 
appearing and disappearing in time, but also the trace of the separated shear layer emerges 
from the presence of small-scale, time-filling events, 
similarly to the region of the upstream boundary layer.
Finally, figure~\ref{fig:spettri_097} shows the wavelet map at $x^*=0.97$, 
the re-attachment location. Large-scale events are now more sparse, indicating that 
the effect of the shock unsteadiness is weaker and a new attached boundary layer is developing.

The scalograms shown above reveal the presence of different events, or structures, at various scales, 
and allows us to detect their degree of intermittency.
In this context, we define the intermittency as the  
alternation of regimes with normal spectral content 
and regimes with significant excess -- or burst -- of energy in a given range of scales.
Given the wavelet transform coefficients $G_{\Psi}(k,\tau)$, it is possible 
to obtain the scale-time distribution of the energy density $|G_{\Psi}(k,\tau)|^2$ of 
the wall-pressure signal $p(t)$ at specified scale $k$ and time $\tau$.
Taking into account this property, \citet{meneveau1991analysis} and \citet{camussi2021intermittent} suggested that an 
effective indicator of the intermittency is the squared local intermittency measure, denoted as $LIM2$:
\begin{equation}
LIM2(k,\tau)=\frac{|G_{\Psi}(k,\tau)|^4}{\langle|G_{\Psi}(k,\tau)|^2\rangle_{\tau}^2}\,. 
\label{eq:LIM2}
\end{equation}
where $\langle \bullet \rangle_\tau$ indicates the time average of the considered quantity.
The rationale for this statement lies in the fact that  
$LIM2$ can be interpreted as a time-scale dependent measure of the flatness factor or kurtosis 
of the input signal $g(t)$, which indicates the importance of rare events for the probability 
distribution of the variable and is defined as the Pearson's index:
\begin{equation}
FF=\frac{\langle g^{'~4} \rangle}{\langle g^{'2}\rangle^2} \,,
\label{eq:FF}
\end{equation}
where $g'$ is the fluctuating part of $g$, and $\langle \bullet \rangle$ indicates the 
expected value of the considered quantity. 
Therefore, the $LIM2$ parameter will be equal to 3 
when the probability distribution is Gaussian, while the condition $LIM2>3$ identifies 
only those rare outliers contributing to the deviation of the wavelet coefficients
from a normal, Gaussian distribution. 
According to \citet{camussi2021intermittent}, 
the energy increment at a certain scale $k$ and time $\tau$ is 
associated with the passage of a coherent structure characterised by the scale $k$.

Figure~\ref{fig:lim_1} (bottom left) shows the $LIM2$ field for the signal at 
$x^*=-0.79$, which has a general kurtosis (see equation \ref{eq:FF}) 
equal to 3.20. 
Only the levels greater than the threshold value 3 are shown. 
In this picture, one can observe the occurrence of the intermittent events 
in the time-scale region characterising the shock unsteadiness (indicated in gray in the background), 
which is the range of large temporal scales defining the broad low-frequency peak 
in the Fourier spectrum (see figure~\ref{fig:fft_shock}). 
Now the significant energy bursts appear more sparse in time 
with respect to the representation offered 
in figure~\ref{fig:spettri_079}. 
In addition, we can also see the intermittency of the turbulent content 
at smaller temporal scales. 
Moreover, we notice that the rather long period considered for the simulation we carried out
is able to capture only few coherent intermittent events at large time scales. For this reason, 
it is evident that, to characterise thoroughly the unsteadiness of \gls{stbli},
very long periods must be considered, in order to deconstruct the broadband low-frequency unsteadiness 
reported in the literature into the actual scattered dynamics associated to the flow.\\
A sign of the presence of intermittent events is already evident from the \gls{pdf} 
of the signal (top right), 
which shows a distribution slightly departing from a normal Gaussian function in 
its external parts. However, although classical statistical indicators  
suggest the relevance of outlier events with large deviation from the mean, 
as indicated by the mildly non-gaussian \gls{pdf} and kurtosis of the signal, they are not able to locate
and estimate precisely the single intermittent episodes, 
which thus makes it hard to understand the physical mechanisms 
behind the resulting dynamics.
The $LIM2$ measure instead allows one to spot the instants and the scales
of data that differ significantly from other observations, and thus makes it possible 
to characterise more accurately the dynamics of the flow, which may be hidden in time-averaged analyses.

After the inspection of such a scenario, we propose 
the following procedure to extract the characteristics of the signal 
of interest \citep{consolini2005local}: 
first, we filter the original wall-pressure using equation~\ref{eq:invFourier}, 
considering only the wavelet coefficients of the large scales 
associated with the low-frequency shock unsteadiness 
(gray region in the $LIM2$ contour); 
then, we reconstruct the intermittent signal $p_{w,I}$ only on the basis of the wavelet coefficients 
related to the burst events
by means of the conditioned inverse continuous wavelet transform:
\begin{equation}
p_{w,I}(k_{min}, t)=\frac{1}{C_{\Psi}} \int_{k_{min}}^{\infty} dk \int_{-\infty}^{\infty} G_{\Psi,\,LIM2}(k,\tau)\Psi\left(\frac{\tau-t}{k}\right)d\tau, 
\label{eq:invLIM}
\end{equation} 
where $C_{\Psi}$ is a normalisation constant depending on the chosen wavelet, 
$k_{min}$ is the smallest time scale considered by the large-scale-pass filtering,
and $G_{\Psi,\,LIM2}(k,\tau)$ are the wavelet coefficients related to the 
intermittent events, extracted using the $LIM2$ condition ($LIM2(k,\tau) > 3$).
%
This procedure can be considered as an extension of the one adopted by \citet{poggie1997wavelet},
in which the authors used the extremes in the global wavelet spectrum to filter the signal 
in the space of the wavelet temporal scales.
Figure~\ref{fig:lim_1} (top) shows the original wall-pressure in time at the foot of the reflected shock 
(green line) together with its filtered-and-reconstructed intermittent part (black line). 
The inspection of the reconstructed signal allows us 
to appreciate that the small-scale turbulent fluctuations have been removed by the procedure, and  
that only the pattern given by the events associated with the shock intermittency remains.
Therefore, it is possible to detect the signal shape 
given by the movement of the shock, which is governed by 
a predominantly two-dimensional mechanism (high span-wise coherence) according to 
\citet{sasaki2021causality}.

The $LIM2$ map can be averaged in time to recover the flatness factor 
as a function of the temporal scale \citep{camussi2021intermittent}, 
and thus to reveal the most intermittent temporal scales.
This quantity is usually called \gls{wff} and is defined as: 
\begin{equation}
 WFF(k)=\langle LIM2(k,\tau)\rangle_{\tau}=\frac{\langle|G_{\Psi}(k,\tau)|^4\rangle_{\tau}}{\langle|G_{\Psi}(k,\tau)|^2\rangle_{\tau}^2}\,.
\label{eq:WFF}
\end{equation}
Figure~\ref{fig:lim_1} (bottom right) shows the \gls{wff} for the corresponding $LIM2$ map  
of the wall-pressure at the shock foot. 
We can see that the \gls{wff} is characterised by two peaks at large scales, 
one at $k\,U_\infty/L_{int} \approx$ 15 and the other at $k\,U_\infty/L_{int} \approx$ 36. 
However, the peak frequency computed by the Fourier and wavelet spectra 
lays in between the scales indicated by the WWF, at $S_L = 0.04$ or $k\,U_\infty/L_{int} = 24.27$. 
This result represents an indication that the most probable temporal scales of intermittency 
are different from the most probable temporal scales of the overall unsteadiness.

\subsection{Wavelet intermittency analysis of the separation bubble dynamics}

\begin{figure*}
 \centering
 \includegraphics[width = \textwidth]{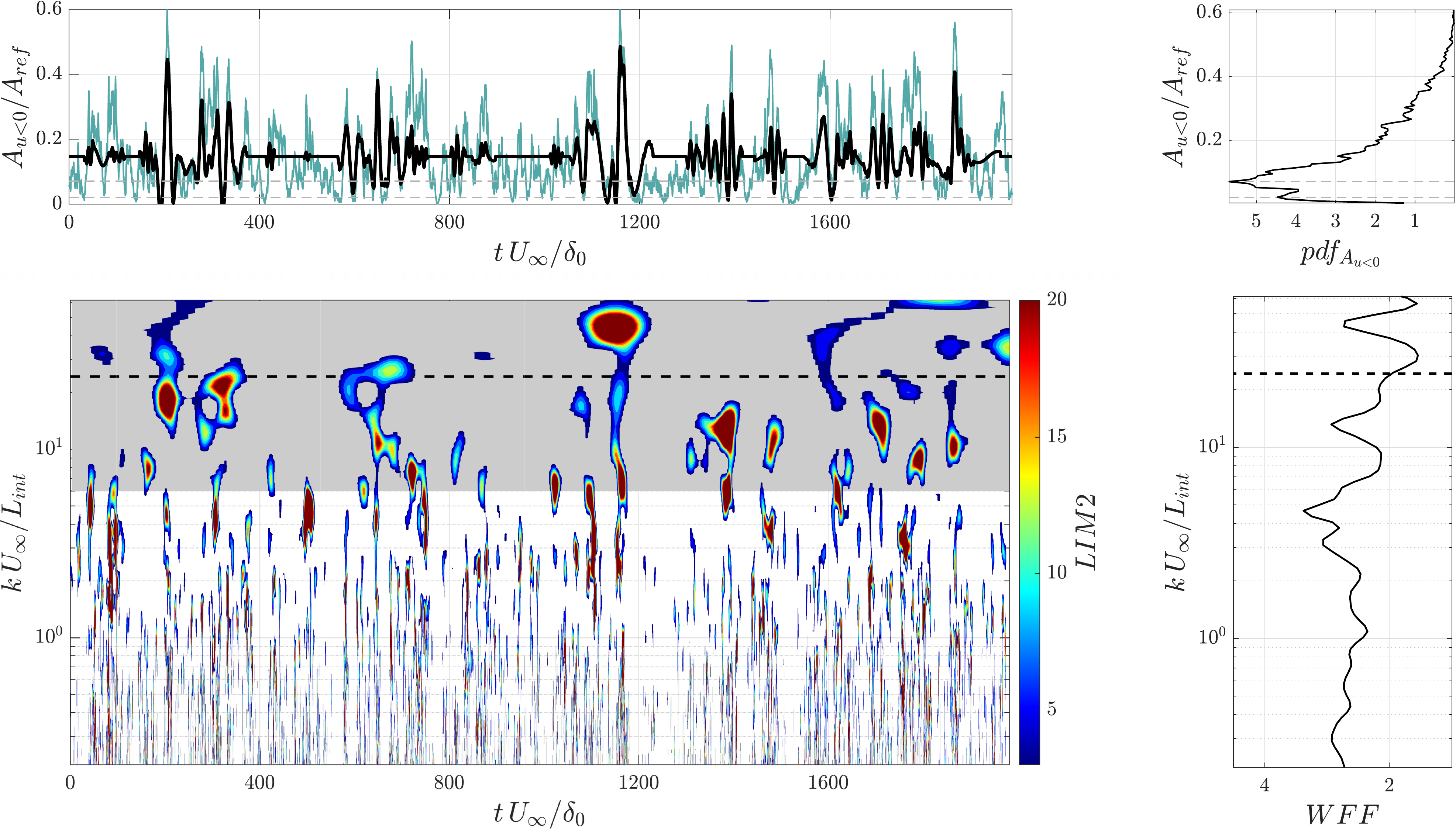} 
 \caption{
            On the top left, the {\color{mygreen}\textbf{green}} curve reports the original 
           $A_{u<0}/A_{ref}$ signal, 
           while the \textbf{black} curve reports its filtered-and-reconstructed part. 
           On the top right, the pdf of the original signal.
           The two peaks of the distribution are indicated with grey dotted lines.
           On the bottom left, a contour of $LIM2$ of $A_{u<0}/A_{ref}$. 
           Values below 3 are cut off, whereas the gray area indicates the scales used for the filtering procedure.  
           On the bottom right, the corresponding WFF.
}
  \label{fig:lim_area}
\end{figure*}

\begin{figure*}
 \centering
 \subfloat[$t \,  U_\infty / \delta_0 = 1158.1$]{
 \includegraphics[width=0.495\textwidth]{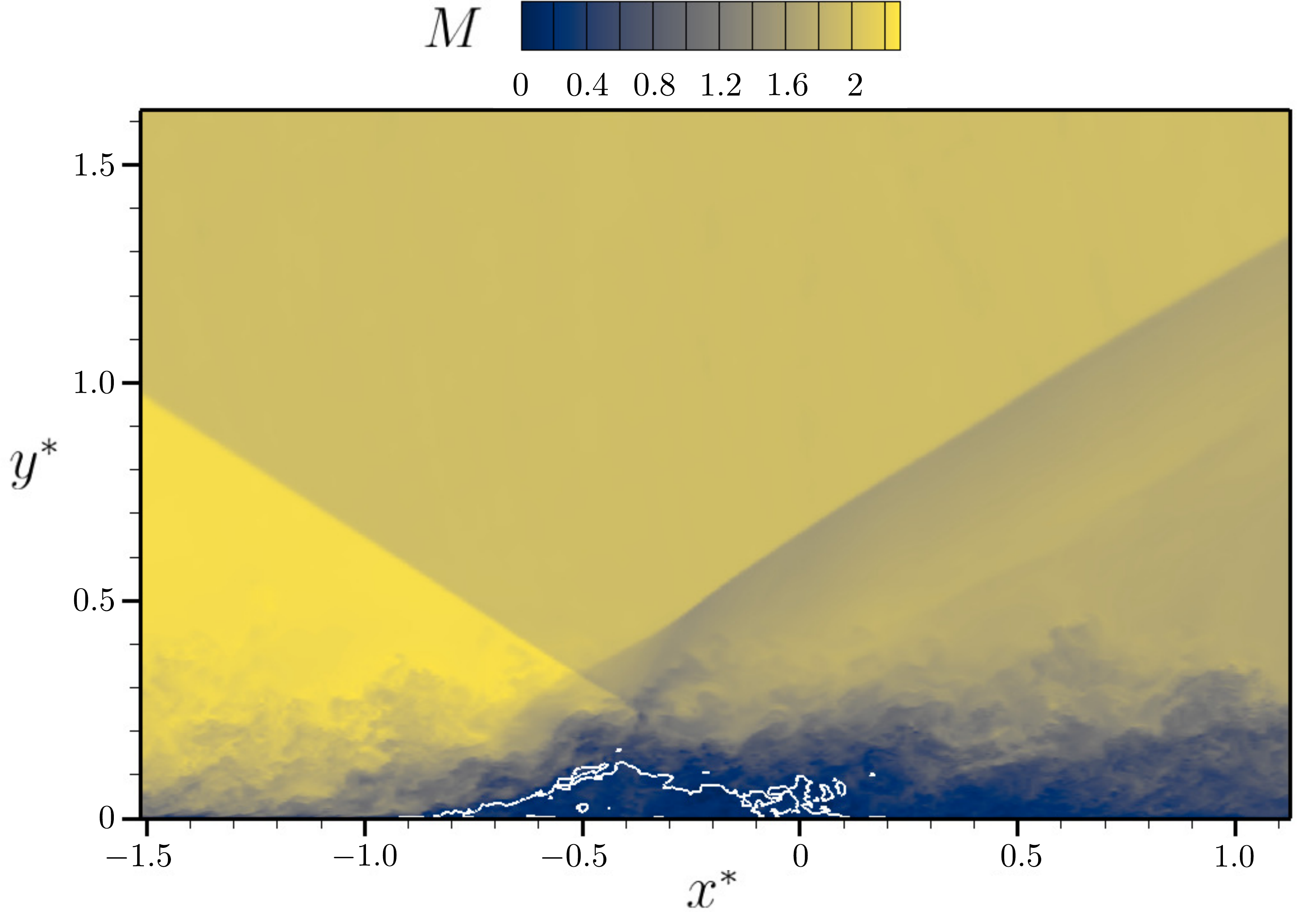}} 
 \subfloat[$t \,  U_\infty / \delta_0 = 1195.6$]{
 \includegraphics[width=0.495\textwidth]{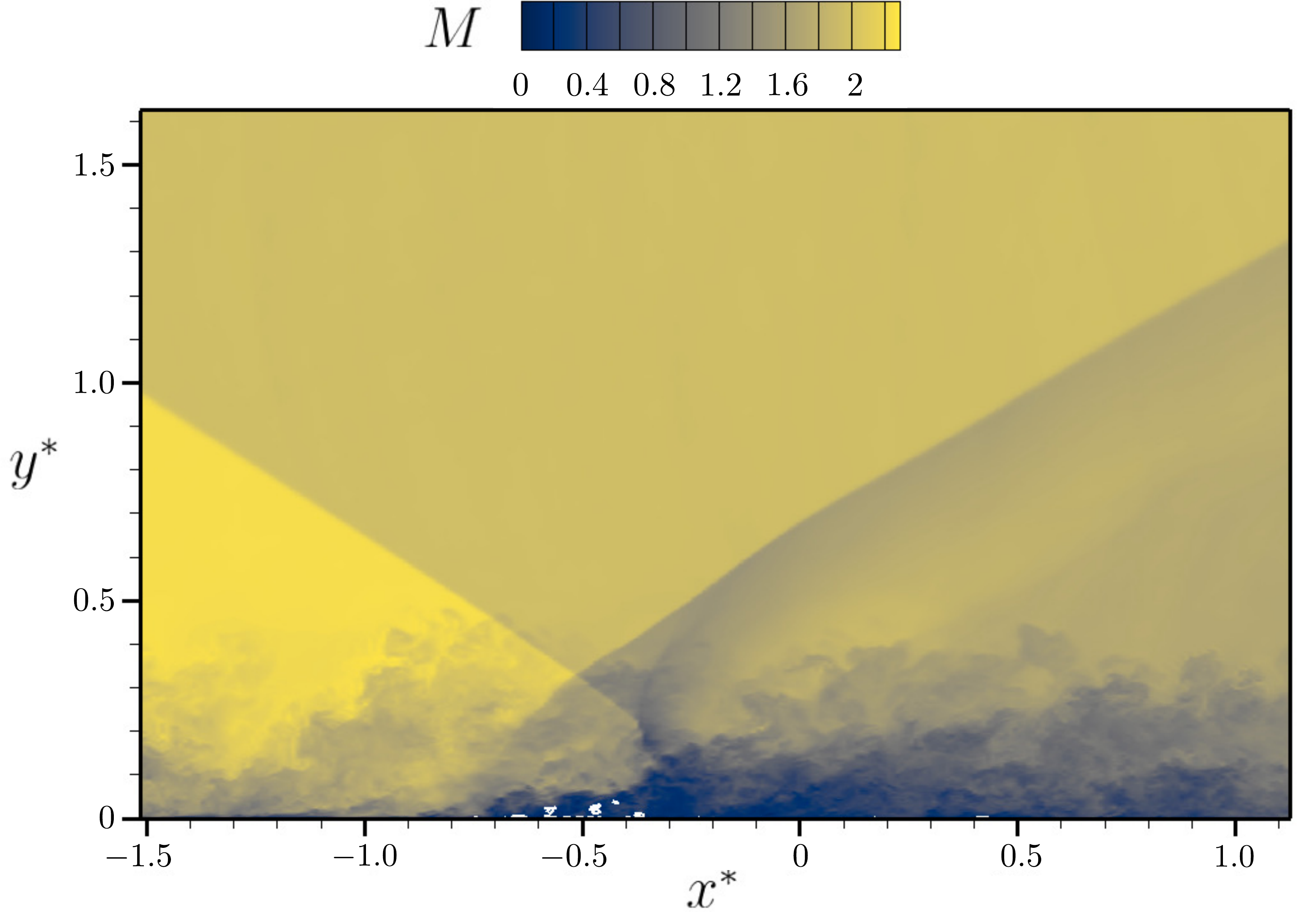}}
 \caption{Instantaneous contours of the Mach number on a vertical slice for two different states during the intermittent event around $t  \,  U_\infty / \delta_0 \approx  1160$. 
  The white line indicates the points with null streamwise velocity. 
  On the left, the area occupied by the bubble is large; on the right, the bubble is almost absent.}
 \label{fig:comparison_mach_separation}
\end{figure*}

\begin{figure*}
     \centering
     \includegraphics[width=0.9\textwidth]{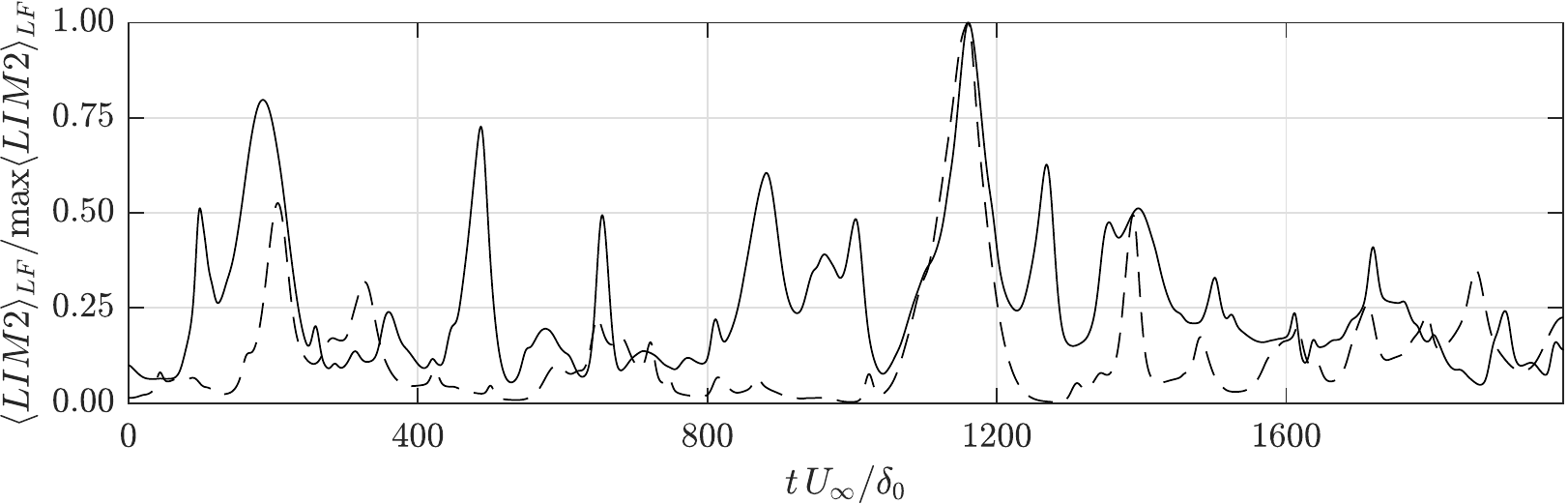} 
     \caption{Normalised scale-average of the $LIM2$ measure above threshold value 3, 
     considering only the large scales -- low frequencies -- used for the filtering procedure.\\
     -- Wall-pressure signal at the shock foot, - - Separation bubble area signal.}
     \label{fig:scale_average}
\end{figure*}

In view of the many works sustaining that the low-frequency shock motion is strictly related to the dynamics of the separation bubble \citep{dupont2006space, piponniau2009simple, pasquariello2017unsteady}, 
we then attempt to relate the behaviour of the wall-pressure at the foot of the shock
to the change in time of the separation region, to assess the eventual correspondence between these two features
and in the attempt of providing a characterisation of their link locally in time and time scale.
In fact, \cite{piponniau2009simple} highlighted that, far from weak interactions with incipient 
separation, where there seems to be an actual correlation \citep{humble2007investigation}, and at least for the case of shock 
reflections, there is no relevant coherence between the upstream superstructures 
and the dynamic response of the system as suggested by \cite{ganapathisubramani2007effects}.
Therefore, for the case under consideration, the dynamics of the separated bubble seems 
clearly the main source of the low-frequency unsteadiness.\\
In order to detect the zone occupied by the recirculating flow,
we recorded the whole area occupied by negative streamwise 
velocity at each instant, 
assuming that this provides a rough estimate of the extent of the bubble. 
The time signal of such area, indicated with $A_{u<0}/A_{ref}$, is reported in figure 
\ref{fig:lim_area} (top) together with its corresponding filtered-and-reconstructed part. 
On the right, we report also the \gls{pdf} of the original signal which has a kurtosis equal to 3.68. 
The extent of the separation bubble oscillates between two more probable values
but presents also many sudden variations at larger areas. The probability density function 
reports thus a highly skewed, bimodal distribution with a long tail that reveals the presence of important 
and intermittent fluctuations in the breathing motion of the separation area.
Contours of the Mach number for two different states during an intermittent event are also shown 
in figure \ref{fig:comparison_mach_separation}, in which the points with null streamwise velocity
are indicated with a white line that gives a hint of the area $A_{u<0}$. 
In the first state, the flow is largely separated, and the deviation imposed on the 
flow by the presence of the separation bubble is significant. 
On the other hand, in the second state, the flow is almost 
completely attached, and the deviation is practically negligible.\footnote{
A movie of the time evolution of the contour of the Mach number on
a vertical slice during the intermittent event around 
$t\,U_\infty/\delta_0 \approx 1160$
is provided in the supplementary material.}
In order to analyse and compare the intermittency of the wall-pressure induced by the oscillation 
of the shock and of the fluctuations of the separation extent, 
we report a contour of the $LIM2$ measure also for $A_{u<0}$ (figure \ref{fig:lim_area}, bottom). 
From the comparison with figure \ref{fig:lim_1}, 
we can observe that some of the most important events of the $LIM2$ of the 
wall-pressure at the shock foot correspond to the most important events of the corresponding measure 
of the separation bubble (see for example $t U_\infty / \delta_0 \approx 200$, $1160$, and $1380$), even though the temporal scales are different. 
The situation is even more evident when we look at Figure \ref{fig:scale_average}, which reports the time evolution of the $LIM2$ intermittency measure averaged in scale, considering only the large scale interval that characterises the low-frequency unsteadiness. 
The intermittency of the pulsation of the separation bubble is thus transmitted to the 
unsteadiness of the shock system, and thus to its wall-pressure signature at the wall, or vice versa. 
However, although many of the intermittent events of the wall-pressure have a corresponding
trace in the $LIM2$ measure of the separation bubble area, other energy bursts 
for the wall-pressure do not seem to have a direct relation 
with the breathing mode of the system, \textit{e.g.} $t U_\infty / \delta_0 \approx 485$, $887$, or $1270$.
This observation suggests that the low-frequency, intermittent unsteadiness of the shock 
can only be in part related to the breathing motion of the separation bubble, and so that
other key actors may play a role in this phenomenon.\\
Finally, if we observe also the $WFF$ of the area signal (figure \ref{fig:lim_area}, bottom right), 
we can see that the distribution of the flatness factor along the time scales follows closely the 
behaviour of the $WFF$ of the wall-pressure (figure \ref{fig:lim_1}, bottom right).
As a result, also for the dynamics of the separation bubble extent,
the most probable intermittent scales are different from 
the large temporal scales characterising the overall shock unsteadiness 
(the broad low-frequency peak in the Fourier domain). 
%
%
%

\begin{figure*}
 \centering
 \subfloat[Original signals. \label{fig:wavelet_coherence_a}]{
 \includegraphics[width=0.485\textwidth]{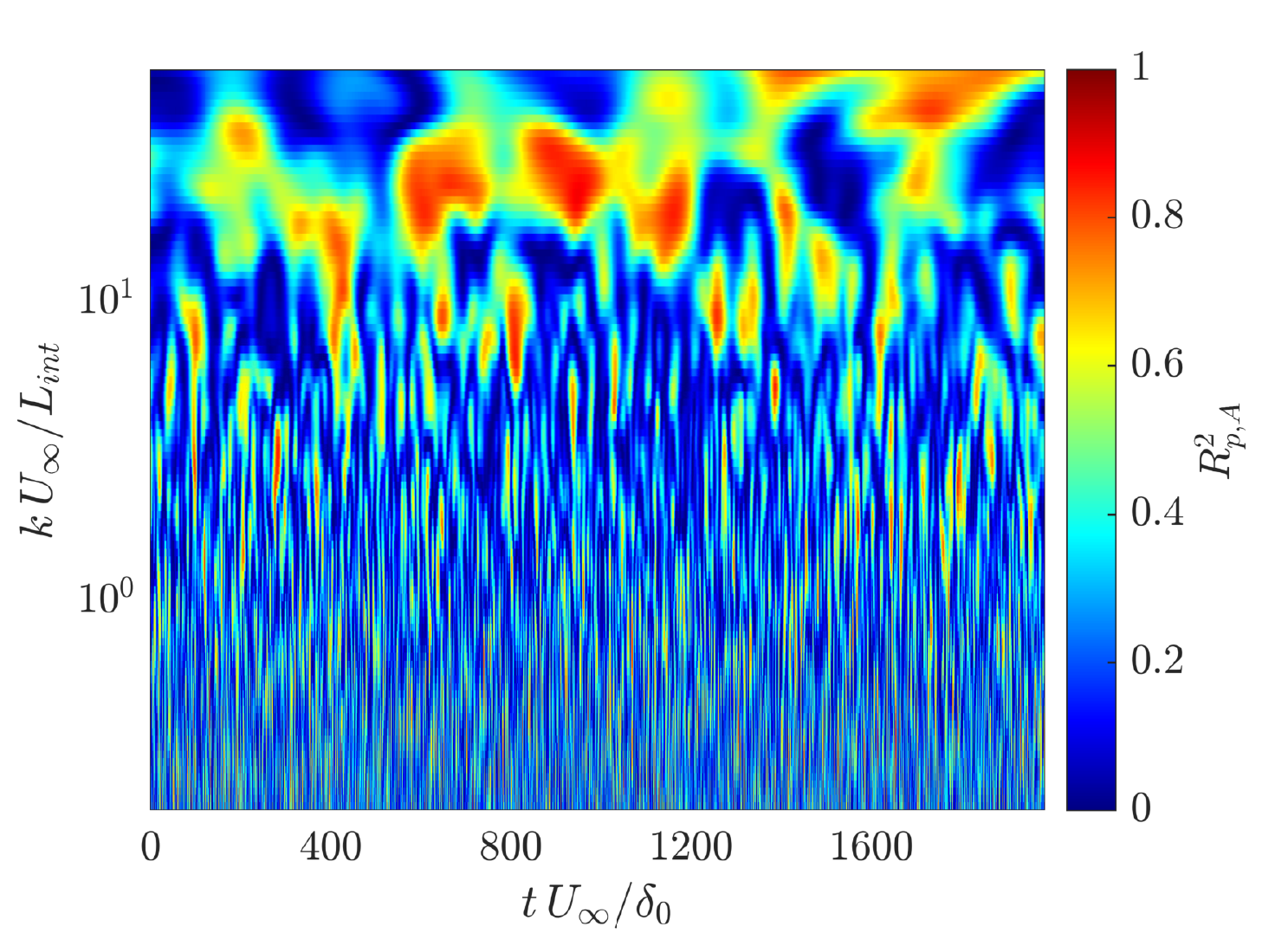}}
 \hfill
 \subfloat[Intermittent signals. \label{fig:wavelet_coherence_b}]{
 \includegraphics[width=0.485\textwidth]{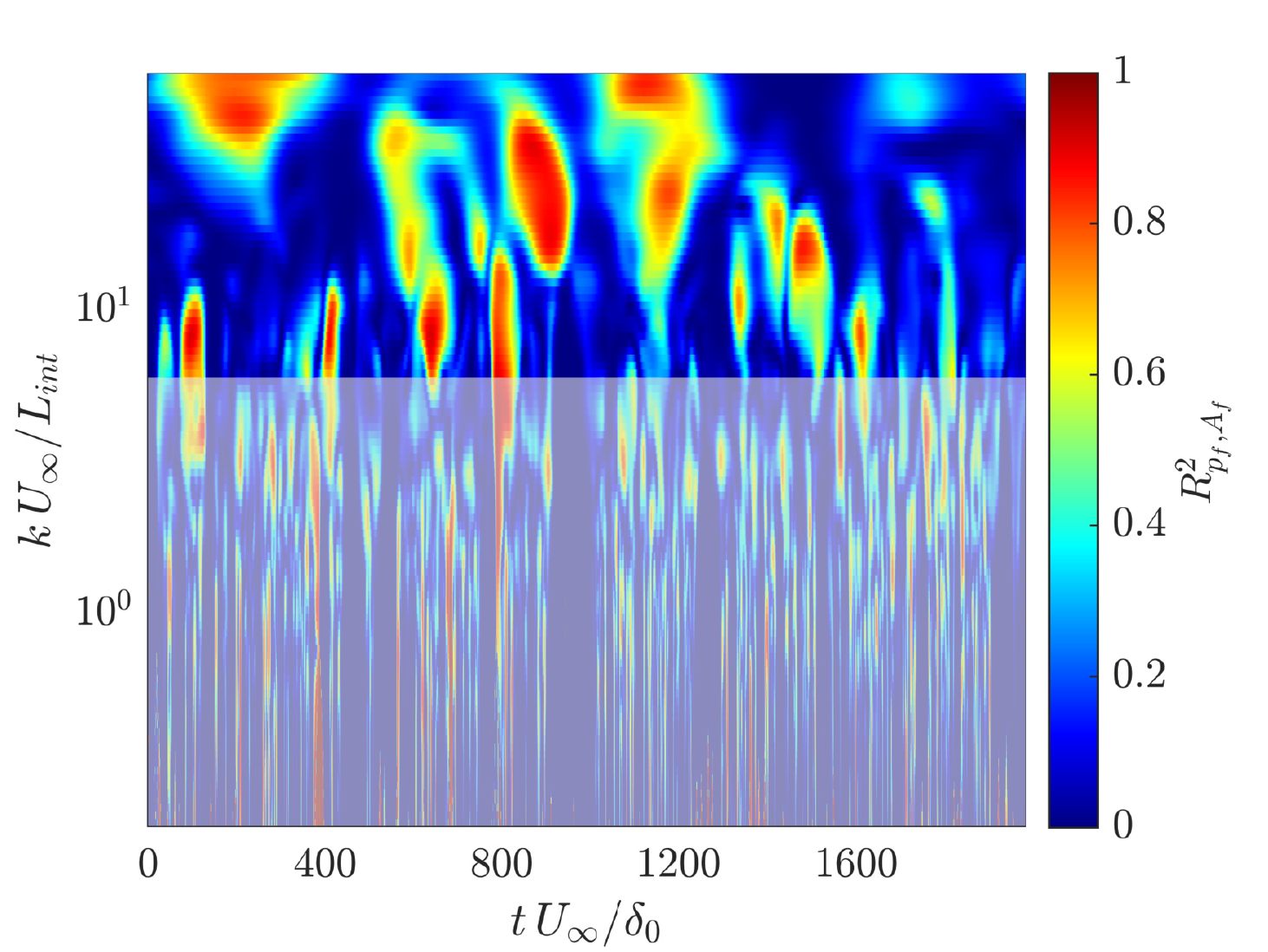}}
 \caption{Wavelet coherence between the shock foot wall-pressure signal and the separation area.
 Arrows indicate the relative phase from the wavelet cross spectrum for the most coherent regions. 
 Shaded region in the right contour is not significant given the preliminary large-scale-pass filtering.}
 \label{fig:wavelet_coherence}
\end{figure*}

\begin{figure*}
 \centering
 \includegraphics[width=0.55\textwidth]{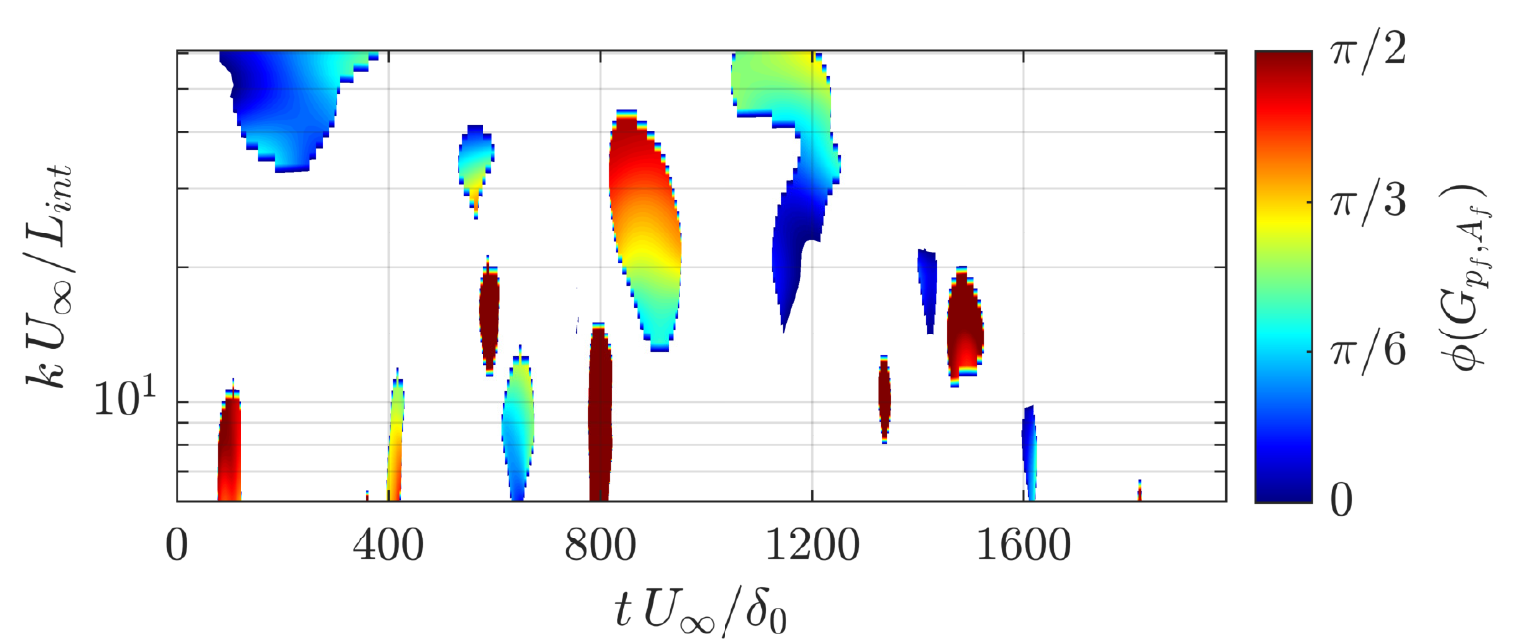}
 \caption{Cross wavelet phase angle for times and time scales for which the wavelet coherence between 
 the filtered wall-pressure and separation area is larger than 0.6.
}
  \label{fig:phase_cross_spec}
\end{figure*}

In order to further characterise the relationship between the signals of the wall-pressure at the 
foot of the shock and the separation area, we consider also the information from the wavelet cross spectrum 
and from the wavelet coherence \citep{torrence1999interdecadal, grinsted2004application, maraun2007nonstationary}. 
Through such analyses, we are able to determine whether there are regions in 
the time/time-scale domain with large common power and also if these have consistent phase relationship, 
which could potentially suggest the causality between the two time series. 
In particular, the wavelet cross spectrum of two time series $x$ and $y$, 
with wavelet transform $G_\Psi^{x}$ and $G_\Psi^{y}$ respectively, is defined as:
\begin{equation}
    G_{x,y} (k, \tau) = G_\Psi^{x} \left(G_\Psi^{y}\right)^*.
\end{equation}
with $*$ indicating the complex conjugation. The wavelet cross spectrum represents thus the covarying power of the two processes. 
This means that an independent variance appearing only in one of the two single spectrum 
does not appear in the cross spectrum, and that the cross spectrum 
vanishes for independent processes. Moreover, for the related time and time scales, the complex argument 
of the wavelet cross spectrum $\phi(G_{x,y})$ gives an indication of the local relative phase between the 
original time series in the time/time-scale domain. 
For example, to establish a simple cause and effect 
relationship between two phenomena represented by the two time series, oscillations are expected to be locked in phase in the 
regions with significant common power \citep{grinsted2004application}. Moreover, starting from the definition of the wavelet cross spectrum, 
it is possible to define a direct measure of the correlation between two signals throughout the time and the scales. 
\citet{torrence1999interdecadal} defined the so-called wavelet coherence $R^2_{x,y}$ between two time series $x$ and $y$ as:
\begin{equation}
    R^2_{x,y} (k, \tau) = \frac{|S(G_{x,y})|^2}{S(|G_\Psi^x|^2) S(|G_\Psi^y|^2)}\,,
\end{equation}
where $S$ is a smoothing operator in time and time-scale. The wavelet coherence has values between zero and one, 
and has the relevant property of quantifying the linear relationship between two processes \citep{maraun2007nonstationary}.
Figure \ref{fig:wavelet_coherence} shows the wavelet coherence between the signal of the wall-pressure at the foot of the 
shock and that of the separation area. In the first instance, we consider the original signals (figure \ref{fig:wavelet_coherence_a}). 
At small time scales, the turbulent fluctuations results in the typical pattern with long, 
intermittent, vertical strikes covering the high-frequency range. 
The coherence is here mostly related to the properties of the small turbulent coherent structures.
In the large-scale -- low-frequency -- range, large areas of the contour show that the two signals present a significant 
instantaneous correlation, although they are not phase locked. 
However, the distribution of the peak regions seems to contradict the correspondence 
of the intermittent events inferred from the comparison of the $LIM2$ measure of the two signals (figure \ref{fig:lim_1} - figure \ref{fig:lim_area}). 
For this reason, we consider also the wavelet coherence of the two signals filtered and reconstructed according to 
the procedure proposed above (figure \ref{fig:wavelet_coherence_b}). Given the large-scale-pass preliminary filtering 
of the signal, the small time-scale region is not considered. From the figure, we can observe an important coherence 
with strong events sparse in time, and most of all, we can recognise the trace of the most important intermittent 
events pointed by the $LIM2$ comparison previously discussed. The signal processing isolates the intermittent content of the 
signal, and the wavelet coherence analysis allows us to reveal that the intermittency of the wall-pressure at the 
foot of the shock is strictly related to the pulsation of the separation area. However, if we look at the phase angle of 
the cross wavelet between the filtered-and-reconstructed signals in figure \ref{fig:phase_cross_spec} for the 
most coherent events, we can see that also for the intermittent signals the phase does not remain constant across the 
time scales, indicating that a simple linear relationship between the intermittency of the breathing mode 
and that of the unsteadiness of the shock can not be inferred. 
%
%
%
\section{Conclusions} \label{conclusions}
The present work analyses the wall-pressure fluctuations of the \gls{stbli}
generated by an impinging oblique shock wave, 
by leveraging the enhanced capabilities of the wavelet analysis 
in the time/time-scale domain \citep{mallat1999wavelet}.

The data are obtained numerically by means of a \gls{dns} carried out at moderate Reynolds number, so that 
the shock is able to penetrate into the 
turbulent boundary layer, leaving a marked trace in the wall-pressure signal
that makes it possible to examine the shock unsteadiness 
and its intermittency more adequately than in the cases with lower Reynolds 
number \citep{Pirozzoli2011}.

Although the 
classical Fourier analysis provides useful information
about the spectral content of the overall \gls{stbli} unsteadiness, 
other techniques may be better able to analyse the transient phenomena that are taking place.
Among them, given its compact support, the wavelet transform represents 
a valuable tool that is still scarcely used in flows similar to the one considered here.
From the analysis of the wall-pressure signals and of the time evolution of the separation bubble, 
we attest in this work
the ability of wavelet analysis to infer relevant information about the dynamics of \gls{stbli}. 

In the first place, 
the validation of our results in terms of basic 
statistics of the fluid field and of the wall-pressure 
shows a notable improvement in the reproduction 
of the reference experiment with respect to previous works. 
Furthermore, the analysis of the wavelet transform modulus
reveals and clarifies
that the broadband low-frequency peak, usually observed in the Fourier spectrum 
at the foot of the reflected shock in \glspl{stbli} \citep{clemens2014low}, 
is actually the result of a collection of sparse, intermittent events 
with different temporal scales. 
The Fourier spectral analysis shows thus only the time-averaged effect 
of such behaviour and is not able to characterise the 
local modulation of the pressure fluctuation, which makes it impossible to 
understand precisely what is happening in the fluid flow correspondingly. 

Then, we show that it is possible to characterise the energy and timing of 
the intermittent events linked to the shock unsteadiness 
on the basis of a wavelet-based local flatness factor, 
called $LIM2$ \citep{meneveau1991analysis, camussi2021intermittent}.
The global kurtosis of the wall-pressure, related to the outliers in the shock signature, 
is thus decomposed into its evolution in the time/time-scale domain, to localise 
the most significant bursts of energy in the signal. 
Inspired by the work of \citet{poggie1997wavelet}, 
the identification of the frequency content characterising 
the shock unsteadiness and the application of the 
$LIM2$ condition have thus been used to propose a 
signal processing procedure to, first, 
separate the footprint of the shock motion from the turbulent 
pressure fluctuations, and then, isolate its intermittent content. 
The analysis of the obtained signal reveals that the intermittency of the wall-pressure 
is mostly characterised by temporal scales that are different from those that 
dominate the overall unsteadiness of the shock system according to Fourier analysis. 

The proposed signal processing can represent a preliminary practice
for further investigations into the causality and the features of
the low-frequency unsteadiness of \gls{stbli} systems, 
by using for example conditional statistics based on the system intermittency. 
In this kind of studies, an essential step is in fact the proper estimation of the 
correlation between the motion of the reflected shock and the characteristic time scales 
of the incoming turbulent boundary layer, the separated region and the re-attachment region:
isolating the low-frequency intermittent part of the associated time series 
could help recognising relations that may be overshadowed  
by the combination of the different components of the complete signals.

Motivated by the many studies in the literature attesting a relationship 
between the unsteadiness of the shock and the pulsation of the separation bubble
\citep{piponniau2009simple}, 
we decided to study by means of the proposed procedure the 
dynamics of the recirculation region behind the reflected shock.
The time signal recording the evolution of the separation bubble extent
presents a skewed probability distribution with 
strong fluctuations between states with large and minimal separation respectively. 
The analysis of the intermittency through wavelet transform 
shows that also the motion of the bubble has strong, sparse energy bursts 
at large time scales and reveals a connection between such events and 
those highlighted by the $LIM2$ measure of the shock foot wall-pressure, 
although it was not possible to estimate the statistics of such events
because of the insufficient number of occurrences. 
In order to qualify and quantify the connection between the two 
signals, wavelet cross spectra and wavelet coherence \citep{torrence1999interdecadal}
have been considered for the signals processed by the proposed procedure. 
The study of the coherence among the times and time scales 
confirms that the intermittent contributions of the breathing bubble motion  
and the shock foot wall-pressure are highly correlated during the 
most important intermittent events. However, the corresponding cross wavelet phase angle
shows that there is not a simple linear relationship between the two signals, 
which would have suggested a causality relation similarly to what recently considered by 
\cite{sasaki2021causality}. 

In the end, our results confirm the hypothesis about the link between the 
low-frequency shock unsteadiness and the breathing mode of the separation bubble, 
even if our investigation points out the complexity of the breathing motion
and indicates that other phenomena may contribute to the dynamics of the overall system.

\backsection[Acknowledgements]{We acknowledge CINECA for the availability of high performance computing resources and support during the pre-production stage of Marconi100 cluster. We also acknowledge Prof. S. Piponniau and Prof. P. Dupont  
for having provided us with the experimental data used in the validation section.}

\backsection[Supplementary data]{\label{SupMat}A movie of the time evolution of a contour of the Mach number on a vertical slice, during the intermittent event around $t \, U_\infty / \delta_0 \approx 1160$ is available. 
The white line indicates the locus of the points with null streamline velocity.}


\backsection[Funding]{This research received no specific grant from any funding agency, commercial or not-for-profit sectors.}

\backsection[Declaration of interests]{The authors report no conflict of interest.}

\backsection[Data availability statement]{The full data set of the DNS simulations is on the order of thousands of gigabytes. By contacting the authors, a smaller subset can be made available.}

\backsection[Author ORCID]{M. Bernardini, \url{https://orcid.org/
0000-0001-5975-3734}; G. Della Posta, \url{https://orcid.org/
0000-0001-5516-9338}; F. Salvadore, \url{https://orcid.org/
0000-0002-1829-3388}; E. Martelli, \url{https://orcid.org/
0000-0002-7681-3513}}


\bibliographystyle{jfm}
\bibliography{main}

\end{document}